\documentclass[12pt, draftclsnofoot, onecolumn]{IEEEtran}

\usepackage{amsmath,epsfig,amssymb,verbatim}
\usepackage{cite}
\usepackage{graphicx}
\usepackage{bm,bbm}
\usepackage{subfigure}

\newtheorem{Theorem}{Theorem}
\newtheorem{Lemma}{Lemma}

\newtheorem{Corollary}{Corollary}

\newcommand{\E}{\mathbb{E}}

\begin{document}

\title{Age of Information of Multi-user Mobile Edge Computing Systems}
\author{Zhifeng Tang,~\IEEEmembership{Student Member,~IEEE,}
Zhuo Sun,~\IEEEmembership{Member,~IEEE,}
Nan Yang,~\IEEEmembership{Senior Member,~IEEE,}\\
and Xiangyun Zhou,~\IEEEmembership{Senior Member,~IEEE}\vspace{-2em}
\thanks{This work was funded by the Australian Research Council Discovery Project (DP180104062).}
\thanks{Z. Tang, N. Yang, and X. Zhou are with the School of Engineering, Australian National University, Canberra, ACT 2600, Australia (Email: \{zhifeng.tang, nan.yang, xiangyun.zhou\}@anu.edu.au).

Z. Sun is with the School of Computer Science, Northwestern Polytechnical University, Xi’an, Shaanxi 710072, China (Email: zsun@nwpu.edu.cn).}
\thanks{Parts of this paper have been published in 2021 IEEE Global Communications Conference \cite{Tang2021globecom}.}

}

\maketitle

\begin{abstract}
In this paper, we analyze the average age of information (AoI) and the average peak AoI (PAoI) of a multiuser mobile edge computing (MEC) system where a base station (BS) generates and transmits computation-intensive packets to user equipments (UEs). In this MEC system, we focus on three computing schemes: (i) The local computing scheme where all computational tasks are computed by the local server at the UE, (ii) The edge computing scheme where all computational tasks are computed by the edge server at the BS, and (iii) The partial computing scheme where computational tasks are partially allocated at the edge server and the rest are computed by the local server. Considering exponentially distributed transmission time and computation time and adopting the first come first serve (FCFS) queuing policy, we derive closed-form expressions for the average AoI and average PAoI. To address the complexity of the average AoI expression, we derive simple upper and lower bounds on the average AoI, which allow us to explicitly examine the dependence of the optimal offloading decision on the MEC system parameters. Aided by simulation results, we verify our analysis and illustrate the impact of system parameters on the AoI performance.
\end{abstract}


\IEEEpeerreviewmaketitle

\section{Introduction}

In recent years, timely status updates have become significantly critical in many emerging real-time applications, such as intelligent transport systems and factory automation \cite{Simsek2016,Li2019}. In order to fully characterize the freshnes`s of delivered status information, the concept of age of information (AoI) was introduced as a new performance metric \cite{Kaul2011}. Specifically, AoI is defined as the elapsed time since the last successfully received status was generated by the transmitter, which is a time metric that captures both the latency and the freshness of a transmitted status. Since being introduced in \cite{Kaul2011}, the concept of AoI has attracted a wide range of interests. First, the AoI performance has been analyzed in point-to-point systems \cite{Kaul2012,Inoue2019,Costa2016J,WangGC2019}. In particular, \cite{Kaul2012} studied the average AoI under the first-come-first-served (FCFS) queuing policy, where three different queue models were considered, i.e., $M/M/1$, $M/D/1$, and $D/M/1$. The last-come-first-served (LCFS) queuing policy was proposed in \cite{Inoue2019}, which was shown to achieve a lower average AoI than the FCFS queuing policy. The authors of \cite{Costa2016J} and \cite{WangGC2019} characterized the impacts of different buffer sizes and status blocklength on the average AoI and the average peak AoI (PAoI), respectively. Based on these studies, the AoI performance was evaluated in multi-user systems \cite{Yates2019,Tang2020,moltafet2021moment,zhou2021performance,Tang2022Lt}. In \cite{Yates2019}, the average AoI was analyzed in a multi-user system under the FCFS queuing policy. By considering sporadic packet generation rates of users, \cite{Tang2020} proposed a random access based transmission scheme to improve the average AoI performance. 
The authors of \cite{moltafet2021moment} derived the moment generating function (MGF) of the AoI of a multi-source system by using the stochastic hybrid systems (SHS) technique. In addition, with the help of SHS technique, \cite{zhou2021performance} derived the closed-form expressions for the average AoI under three policies in an Internet of things (IoT) system and discussed the AoI performance with considering a sufficient large number of users. 
Furthermore, \cite{Tang2022Lt} designed a Whittle index based scheduling policy to optimize the AoI performance over unreliable channels. 


The aforementioned papers mainly focused on the impact of data transmission on the AoI performance. In fact, data processing, which usually consumes a significant amount of time, also affects the AoI performance. Particularly, when data processing is computationally intensive, the local server with a limited computing capacity results in time-consuming data processing, which seriously degrades the AoI performance. To tackle this problem, mobile edge computing (MEC) was introduced by the European Telecommunications Standard Institute (ETSI) \cite{Hu2015White}. In MEC systems, the server deployed at the network edge, called the edge server, is exploited to partially offload data processing tasks from the local server \cite{Mao2017,Mach2017}. Owing to the powerful computing capacity of the edge server, the MEC system can significantly reduce the computation time. In MEC systems, the main challenge is to determine whether or not to offload and how much should be offloaded \cite{Zhang2012}, which is affected by a number of parameters, such as the computing capacity of the edge server, the number of users, as well as the status transmission rate \cite{Chen2015}. In the literature, \cite{Gong2013,Sardellitti2015,Liu2016,Mao2016,You2017,Rodrigues2018,Zhao2019,Ren2019} investigated the computation offloading to minimize the execution delay in different MEC systems.

The AoI has been widely evaluated as an effective performance metric of MEC systems, starting from point-to-point systems. 
The authors of \cite{Kuang2020} investigated the impact of different offloading schemes on the AoI performance of the MEC system. Also, \cite{Kuang2020} found that partially offloading tasks to the edge server achieves the better AoI performance than other schemes by carefully partitioning tasks. 
By considering different packet management policies, \cite{Zou2021ACM} studied the AoI performance of an MEC system. Moreover, based on the concept of the AoI, \cite{Li2021AgePr} proposed a novel metric, age of processing (AoP), to quantify the freshness of information of the MEC system and designed an offloading policy to improve the AoP performance of the MEC system. The authors of \cite{Arafa2019} designed a cutoff policy for an MEC system, where a fresh packet replaces the previous packet once the computation time of the previous packet is larger than a threshold. 
Furthermore, \cite{Buyukates2022Tcom} investigated the AoI performance of an MEC system with different coding schemes and found that the multiple computations coding scheme minimizes the average AoI.

Building upon these efforts on the MEC system with a single user, increasing research efforts have been devoted to investigating the AoI performance of multi-user MEC systems. For example, \cite{Xu2020IoTJ} derived the closed-form expressions for the average PAoI and designed a min–max optimization problem to minimize the maximum average PAoI of an MEC system. The authors of \cite{Zhou2020J} jointly designed a scheduling and status sampling policy to minimize the average AoI of the MEC system. In addition, \cite{wu2020data} designed the edge resource allocation to minimize the average AoI of a multi-user MEC system. 
  Although the aforementioned studies \cite{Kuang2020,Zou2021ACM,Li2021AgePr,Arafa2019,Buyukates2022Tcom,Xu2020IoTJ,Zhou2020J,wu2020data} have designed the scheduling policy to minimize the average AoI and analyzed the AoI performance of different MEC systems, they have not touched a key problem: ``How many computational tasks need to be offloaded in a multi-user MEC system such that its AoI performance is optimized?'' To solve this problem, the impact of system parameters on the AoI performance needs to be investigated. We highlight that it is not trivial to design offloading in the multi-user MEC system, since in this system the transmission and computation of each user's status is affected by other users’ status. Thus, the design in the single-user MEC system cannot be applied into the multi-user MEC system.

\begin{figure*}[t]
    \normalsize
    \centering
    \includegraphics[width=0.8\columnwidth]{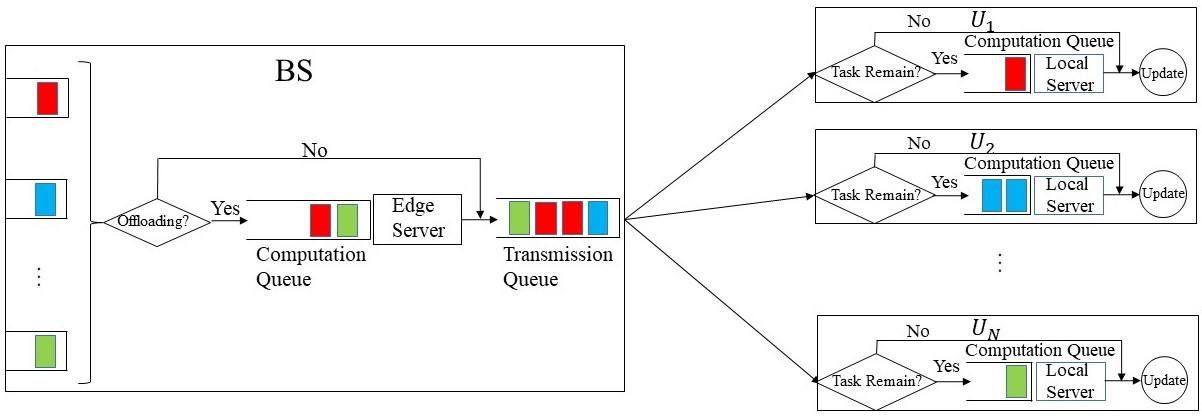}
    \vspace{-2em}
    \centering
    \caption{Illustration of our considered MEC system where the BS transmits computation-intensive packets to $N$ UEs.}\label{fig:system_model}
    \vspace{-2em}
\end{figure*}

In this paper, we analyze the average AoI and the average PAoI of a multi-user downlink MEC system. In this system, the base station (BS) transmits computation-intensive packets to multiple user equipments (UEs). Each packet can be computed by the local server at each UE, referred to as the local computing scheme, or by the edge server at the BS, referred to as the edge computing scheme, or by both of them, referred to as the partial computing scheme. In these three computing schemes, computation and transmission of packets follow the FCFS queuing policy and computation time and transmission time of each packet follow exponential distributions. The main contribution of this paper are summarized as follows:
\begin{itemize}
    \item We derive the closed-form expressions for the average AoI and the average PAoI of three computing schemes. Using simulations, we demonstrate the accuracy of our analysis results. We also find that the edge computing scheme achieves the lowest average AoI compared with the local computing scheme and the partial computing scheme when the packet generation rate and the number of UEs are small. We further find that by carefully partitioning the computational tasks, the partial computing scheme has a better system stability than the other two computing schemes.
    \item We derive the upper bound and the lower bound on the average AoI. With a sporadic packet generation pattern, we find that the average AoI is close to the average PAoI. We further obtain the optimal offload ratio to minimize the average PAoI. 
     Simulations show that the gap between the optimal average AoI and the average AoI with the offloading ratio to minimize the average PAoI is negligible regardless the packet generation rate. This optimal offloading ratio intuitively indicates how much proportion of tasks should be offloaded to minimize the average AoI and PAoI in a multi-user MEC system. 
    \item We investigate the impact of the system parameter on the optimal offloading ratio of the MEC system. We observe that the optimal offloading ratio is proportional to the computation rate of the edge server and inversely proportional to the computation rate of the local server and the number of UEs. We further find that the transmission rate hardly affects the optimal offloading ratio in the partial computing scheme. 
    
    
\end{itemize}

We clarify that this work provides sufficient novel contributions comparing to \cite{Tang2021globecom}. First, we propose a new partial computing scheme and derive its closed-form expressions for the average AoI and average PAoI in the MEC system. Second, to address the complexity of the average AoI expression, we obtain simple expressions for the upper bound and lower bound on the average AoI. Third, we derive an optimal offloading ratio for the partial computing scheme to minimize the average PAoI and find that this offloading ratio can also be adopted to minimize the average AoI in the MEC system. All such contributions were not considered in \cite{Tang2021globecom}.

The rest of the paper is organized as follows. In Section \ref{Sec:systemmodel}, the system model and the average AoI and PAoI of the considered MEC system are presented. The closed-form expressions for the average AoI and PAoI of three schemes are derived in Section \ref{sec:Derivation}. In Section \ref{Sec:Approx}, the upper bound and the lower bound on the average AoI and the optimal offloading ratio are presented. The numerical results are discussed in Section \ref{sec:Numerical}. This paper is concluded in Section \ref{sec:Conclusion}.

\section{System Model and Average AoI}\label{Sec:systemmodel}

In this paper, we consider a downlink system, as depicted in Fig.~\ref{fig:system_model}, where the BS transmits time-sensitive packets to $N$ UEs. We denote the $n$th UE by $U_{n}$, where $n=1,2,\cdots,N$. In this system, the BS generates the packet of $U_n$ according to a Poisson process\footnote{We assume that the packet generation processes among UEs are independent but not identical.} with the rate $\lambda_{n}$\footnote{In practical applications, e.g., a vehicular network, the BS can generate traffic information for vehicles.}. We assume that packets are generated with indices (e.g., using the first several bits of each packet) to associate with UEs and the BS transmits packets to their corresponding UEs based on the indices. To ensure the freshness of packets, we consider an MEC system, where 
a proportion of total computational tasks of a packet can be processed at the local server and the remaining part is processed at the edge server. Such separate processing is reflected in practical applications, such as the virus scan application and the file compression application \cite{Wang2016T}, where the computational tasks of packets are divided into several independent subsets that can be processed individually. In this MEC system, we introduce three computing schemes, namely, the local computing scheme, the edge computing scheme, and the partial computing scheme.

\subsection{Computing Schemes}

Depending on which server computes the computational tasks of packets, we introduce three computing schemes as follows:

\begin{enumerate}
    \item \textit{Local Computing}: In this scheme, the BS directly transmits the generated packet to the UE for local computing. In particular, when a packet of $U_n$ is generated by the BS, denoted by $P_n$, it waits to be transmitted in the transmission queue at the BS. After being transmitted to $U_n$, $P_n$ arrives at the computation queue at the local server at $U_n$. Finally, the local server completes the computation of $P_n$ and $U_n$ obtains the computational result of $P_n$.
    \item \textit{Edge Computing}: In this scheme, the edge server at the BS performs the computation of the packet and transmits the computational result to the UE. In particular, the generated packet $P_n$ arrives at the computation queue at the edge server. After the computation is completed by the edge server, the computational result of $P_n$ arrives at the transmission queue and is transmitted to $U_n$. 
    \item \textit{Partial Computing}: In this scheme, the edge server partially computes the packet and then the BS transmits the intermediate computational result to the UE for the remaining computation. Specifically, the generated packet $P_n$ first arrives at the computation queue at the edge server for a partial computation and the intermediate computational result of $P_n$ arrives at the transmission queue. When $U_n$ receives the intermediate computational result of $P_n$, its local server performs the remaining computation to obtain the computational result of $P_n$. 
    
\end{enumerate}

In these three computing schemes, we assume that the packets in the queues are served by using the FCFS queuing policy and each queue has infinite buffer size to store the packet. We clarify that these $N$ UEs compete with each other in the computation queue of the edge server and the transmission queue at the BS, due to the FCFS queueing policy. This competition among UEs affects the waiting time of the packet in the computation queue of the edge server and the transmission queue at the BS. We then assume that both the computation time and the transmission time of a packet follow exponential distributions, which are commonly used for the AoI analysis in \cite{Yates2019,Sthapit2019,Kuang2020}. On one hand, the packet transmission time depends on wireless channels. It may or may not incorporate retransmission and backoff. It has been shown that such complex effects are accurately characterized by an exponential distributed transmission time in \cite{Yates2019}. On the other hand, the packet computation time depends on the complexity of computational tasks. For example, \cite{Sthapit2019} showed that an exponential distributed computation time is accurate to characterize the different complexity of computational tasks in face recognition. Then, the probability density functions (PDFs) of the computation time of a packet at the edge server, $S_B$, the transmission time of a packet, $S_D$, and the computation time of a packet at the local server at $U_n$, $S_{U_n}$, are given by
\begin{align}
    f_{S_B}(t) = \exp(-\mu_B t),
\end{align}
\begin{align}
    f_{S_D}(t) = \exp(-\mu_D t),
\end{align}
and
\begin{align}
    f_{S_{U_n}}(t) = \exp(-\mu_n t),
\end{align}
respectively, 
where $\mu_B$ is the computation rate of the edge server, $\mu_D$ is the transmission rate, and $\mu_n$ is the computation rate of the local server at $U_n$. Here, $\mu_B$ indicates the average number of packets served per unit time at the edge server in the edge computing scheme, $\mu_n$ indicates the average number of packets served per unit time at the local server at $U_n$ in the local computing scheme, and $\mu_D$ indicates the average number of packets transmitted per unit time from the BS to UEs\footnote{We assume that the information bits of the original packet, the partial computed packet, and the computed packet are same, such that the transmission rate of these packets are same.}.

In order to fairly compare the AoI performance of these three schemes, we assume that these three computing schemes share the same transmission rate and computation rate at each server. For the partial computing scheme, we adopt a linear computation partitioning model introduced in \cite{Wang2016T}. In particular, we denote $p$ as the offloading ratio, which represents the percentage of computational tasks of each packet computed by the edge server, where $0\leq p\leq 1$. Moreover, the effective computation rate of the edge server is denoted by $\mu_B'$ and the effective computation rate of the local server at $U_n$ is denoted by $\mu_n'$. Here, $\mu_B'$ and $\mu_n'$ indicate the average number of packets served per unit time at the edge server and the local server at $U_n$, respectively, which depend on not only the performance of servers but the proportion of tasks that need to be computed in each packet at the server. Thus, we have $\mu_B'=\frac{\mu_B}{p}$ and $\mu_n'=\frac{\mu_n}{1-p}$. We clarify that the local computing scheme is considered as a special case of the partial computing scheme where the offloading ratio is $0$, i.e., $p=0$. In this scheme, the serving time of packets at the edge server is $0$. In addition, the edge computing scheme is considered as another special case of the partial computing scheme where the offloading ratio is $1$, i.e., $p=1$. In this scheme, the serving time of packets at the local server at $U_n$ is $0$.

\subsection{Average AoI and PAoI}

\begin{figure}[t]
    \centering
    \includegraphics[width=0.75\columnwidth]{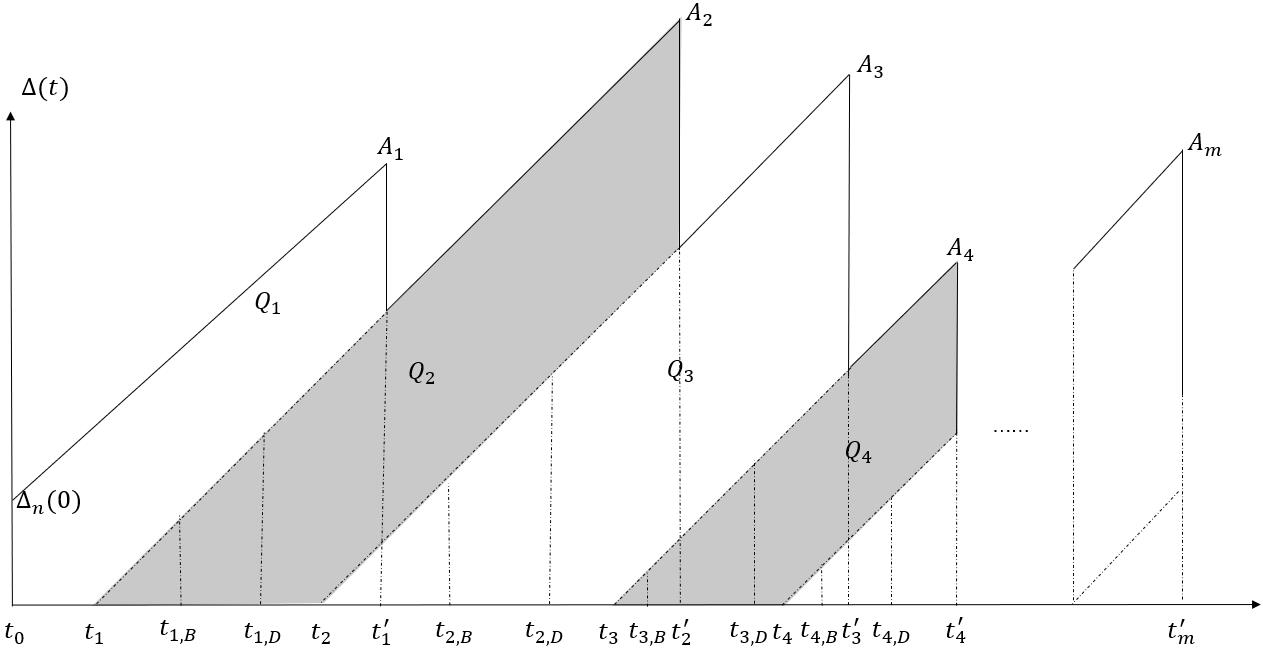}
    \vspace{-2em}
    \caption{The AoI variation of the selected UE, $U_n$.}\label{fig:AoIevo}
    \vspace{-2em}
\end{figure}

In this subsection, we present the general expression for the average AoI and PAoI of the considered MEC system. Recall that our target is to analyze the AoI performance of the considered MEC system. 
Without loss of generality, we arbitrarily select one UE, $U_n$, and analyze its average AoI, $\Delta_n$, and average PAoI, $\Omega_n$. We will define the average AoI and the average PAoI of the system at the end of this section based on the individual UE's AoI. Fig.~\ref{fig:AoIevo} plots a sample variation of AoI for $U_n$, $\Delta_n(t)$, as a function of $t$. We assume that the observation begins at $t=0$, where the AoI is $\Delta_n(0)$. From Fig.~\ref{fig:AoIevo}, we express the AoI of $U_n$ at time $t$ as
\begin{align}
    \Delta_n(t) = t - u_n(t),
\end{align}
where $u_n(t)$ is the generation time of the last received computed packet of $U_n$ at time $t$. Then the time-average AoI of $U_n$ over the observation time interval $(0,\tau)$ can be calculated as
\begin{align}\label{eq:AoIEQ5}
    \Delta_n = \frac{1}{\tau}\int_{0}^{\tau}\Delta_n(t) \mathrm{d}t. 
\end{align}

We denote $P_{n,j}$ as the $j$th packet generated after time $t=0$ of $U_n$, $j=1,2,\cdots$. We then denote $Y_j$ as the time interval from the generation time of $P_{n,j-1}$ to the generation time of $P_{n,j}$ and denote $T_j$ as the time interval from the generation time of $P_{n,j}$ to the time that $U_n$ obtains the computational result of ${P}_{n,j}$. The value of age in the peak is denoted by $A_j$. Therefore, we obtain
\begin{align}
    Y_j &= t_{j}-t_{j-1},
\end{align}
\begin{align}    
    T_j &= t_j'-t_j,
\end{align}
and
\begin{align}    
    A_j &= Y_j+T_j,
\end{align}
where $t_j$ is the generation time of $P_{n,j}$ and $t_j'$ is the time that $U_n$ obtains the computational result of ${P}_{n,j}$. We note that $Y_1=t_1$ is obtained by setting $t_0=0$. We consider that $U_n$ obtains the computational result of the $m$th packet at the end of this time interval, i.e.,  $\tau=t_m'$. We note that $\int_{0}^{\tau}\Delta_n(t)\mathrm{d}t$ in \eqref{eq:AoIEQ5} is the sum of disjoint areas $Q_j$, $j=\{1,2,\cdots,m\}$, shown in Fig.~\ref{fig:AoIevo}, and the triangular area of width $T_m$ over the time interval $(t_m,t_m')$. Hence, we rewrite the average AoI as
\begin{align}\label{eq:aveAoIPar2}
    \Delta_n\! = \!\frac{\sum\limits_{j\!=\!1}^m Q_j\!+\!\frac{T_m^2}{2} }{\tau}\!=\! \frac{2Q_1\!+\!T_m^2}{2\tau}\!+\!\frac{m\!-\!1}{\tau}\!\times\!\left(\!\frac{1}{m\!-\!1}\sum\limits_{j\!=\!2}^m      Q_j\!\right).
\end{align}
From Fig.~\ref{fig:AoIevo}, we see that $Q_1$ is a polygon and $Q_j$ is an isosceles trapezoid for $j\geq 2$, which can be derived from two isosceles triangles, i.e.,
\begin{align}\label{eq:Qjexpression}
    Q_j &= \frac{1}{2}(Y_j+T_j)^2-\frac{1}{2}T_j^2=\frac{Y_j^2}{2}+Y_jT_j.
\end{align}
We note from \eqref{eq:aveAoIPar2} that, when $\tau\rightarrow\infty$, the impact of $Q_1$ and $T_m^2$ on the average AoI is negligible, i.e., $\lim_{\tau\rightarrow\infty}\frac{2Q_1+{T_m^2}}{2\tau} = 0$, because $Q_1$ and $T_m^2$ are finite. Moreover, due to the fact that $\tau=Y_1+\sum_{j=2}^m Y_j+T_m$, we obtain $\lim_{\tau\rightarrow\infty}\frac{\tau}{m-1} =\E[Y_j]$, where $\E[\cdot]$ is the expectation. Therefore, by substituting \eqref{eq:Qjexpression} into \eqref{eq:aveAoIPar2} and taking $m$ to infinity, we obtain the average AoI of $U_n$ as 
\begin{align}\label{eq:newAoIevo1}
    \Delta_n\!=\!\frac{\lim\limits_{m\rightarrow\infty}\frac{1}{m-1}\sum\limits_{j=2}^m      Q_j}{\E[Y_j]} \!= \!\frac{\E[Q_j]}{\E[Y_j]}\!=\! \frac{\E[Y_j^2]\!+\!2\E[Y_jT_{j}]}{2\E[Y_j]}.
\end{align}
We keep the index in \eqref{eq:newAoIevo1} to reflect the correlation between $Y_j$ and $T_j$. As the BS generates the packet of $U_n$ according to a Poisson process with the rate $\lambda_n$, we obtain $\E[Y_j^2] = \frac{2}{\lambda_n^2}$ and $\E[Y_j] = \frac{1}{\lambda_n}$, which leads to
\begin{align}\label{eq:newAoIevo}
    \Delta_n= \frac{1}{\lambda_n}+\lambda_n\E[Y_jT_{j}].
\end{align}
For the average PAoI, we calculate it as
\begin{align}\label{eq:newPAoIevo}
    \Omega_n = \lim_{m\rightarrow \infty}\frac{1}{m}\sum\limits_{j=1}^{m}A_j=\frac{1}{\lambda_n}+\E[T_j].
\end{align}

We denote $t_{j,B}$ as the time that $P_{n,j}$ arrives at the transmission queue and $t_{j,D}$ as the time that $P_{n,j}$ arrives at the computation queue at the local server at $U_n$. We then denote $X_{j,B}$, $X_{j,D}$, and $X_{j,U}$ as the queuing delay of $P_{n,j}$ in the computation queue at the edge server, the queuing delay of $P_{n,j}$ in the transmission queue, and the queuing delay of $P_{n,j}$ in the computation queue at the local server at $U_n$, respectively. They are expressed as 
\begin{align}
    X_{j,B} = t_{j,B} - t_j,\\
    X_{j,D} = t_{j,D} - t_{j,B},
\end{align}
and
\begin{align}
    X_{j,U} = t_j'-t_{j,D},
\end{align}
respectively. Then, we rewrite $T_j$ as
\begin{align}\label{eq:Tj1}
    T_j &= X_{j,B}+X_{j,D}+X_{j,U}.
\end{align}
We now denote $W_{j,B}$, $W_{j,D}$, and $W_{j,U}$ as the waiting time of $P_{n,j}$ in the computation queue at the edge server, the waiting time of $P_{n,j}$ in the transmission queue, and the waiting time of $P_{n,j}$ in the computation queue at the local server at $U_n$, respectively. Next, we denote $S_{j,B}$, $S_{j,D}$, and $S_{j,U}$ as the computation time of $P_{n,j}$ at the edge server, the transmission time of $P_{n,j}$, and the computation time of $P_{n,j}$ at the local server at $U_n$, respectively. Note that the queuing delay of $P_{n,j}$ in each queue is the summation of the waiting time and the serving time of $P_{n,j}$ in this queue. Thus, we rewrite $X_{j,B}$, $X_{j,D}$, and $X_{j,U}$ as
\begin{align}
    X_{j,B} = W_{j,B}+S_{j,B},\label{eq:XjBsec2}\\
    X_{j,D} = W_{j,D}+S_{j,D},\label{eq:XjDsec2}
\end{align}
and
\begin{align}
    X_{j,U} = W_{j,U}+S_{j,U},\label{eq:XjUsec2}
\end{align}
respectively. By substituting \eqref{eq:XjBsec2}, \eqref{eq:XjDsec2}, and \eqref{eq:XjUsec2} into \eqref{eq:Tj1}, we obtain
\begin{align}\label{eq:Tj}
    T_j =W_{j,B}+S_{j,B}+W_{j,D}+S_{j,D}+W_{j,U}+S_{j,U}.
\end{align}
We note that the packet generation is independent of the packet transmission and computation in each queue. By substituting \eqref{eq:Tj} into \eqref{eq:newAoIevo} and \eqref{eq:newPAoIevo}, and using the fact that $S_{j,B}$, $S_{j,D}$, and $S_{j,U}$ are independent of $Y_j$, we obtain 
\begin{align}\label{eq:newAoIevox}
     \Delta_n=& \frac{1}{\lambda_n}+\E[S_{j,B}]+\E[S_{j,D}]+\E[S_{j,U}]+\lambda_n\left(\E[Y_jW_{j,B}]+\E[Y_jW_{j,D}]+\E[Y_jW_{j,U}]\right),
\end{align}
and
\begin{align}\label{eq:newPAoIevox}
     \Omega_n=& \frac{1}{\lambda_n}+\E[S_{j,B}]+\E[S_{j,D}]+\E[S_{j,U}]+\E[W_{j,B}]+\E[W_{j,D}]+\E[W_{j,U}].
\end{align}

By averaging $\Delta_n$ over all UEs, we obtain the average AoI and PAoI of the MEC system as 
\begin{align}\label{eq:AveAoIallUE}
    \Delta = \frac{1}{N} \sum\limits_{n=1}^{N}\Delta_n,
\end{align}
and
\begin{align}\label{eq:AvePAoIallUE}
    \Omega = \frac{1}{N} \sum\limits_{n=1}^{N}\Omega_n,
\end{align}
respectively. It is worthwhile to note that, when one of the expectations of $W_{j,B}$, $W_{j,D}$, and $W_{j,U}$ in \eqref{eq:newAoIevox} becomes infinity, the average AoI and the average PAoI of $U_n$ go to infinity. In this case, the MEC system is considered to be unstable. Thus, in order to ensure system stability, we assume that the arrival rate of the packet is lower than its serving rate in each queue in this paper.

\section{Characterizing Average AoI and PAoI}\label{sec:Derivation}

In this section, we derive the closed-form expressions for the average AoI and PAoI of three computing schemes. We first derive the closed-form expression for the average AoI and PAoI of the partial computing scheme, since both the local computing scheme and the edge computing scheme are special cases of the partial computing scheme. Based on that, we then obtain the closed-form expressions for the average AoI and PAoI of the local computing scheme and the edge computing scheme, respectively.

\begin{figure*}[t]

\begin{align}
&\Phi_{q,B_{j,D}}=\frac{\lambda_n(\mu_B'+\mu_D-\lambda)}{\mu_B'(\mu_D\!-\!\lambda)(\mu_B'\!+\!\mu_D\!-\!\lambda\!-\!\lambda_{-n})^2}+\frac{\lambda_n(\mu_B'-\lambda)(\mu_B'-\lambda_{-n})\left(\frac{1}{(\mu_D-\lambda_{-n})^2}-\frac{1}{(\mu_B'-\lambda_{-n})^2}\right)}{(\mu_B'-\mu_D)(\mu_D-\lambda)(\mu_D+\mu_B'-\lambda-\lambda_{-n})}\notag\\
     &+\frac{\lambda_n\lambda_{-n}(\mu_B'\!-\!\lambda)(\mu_B'\!-\!\lambda_{-n})\left(\frac{2}{(\mu_D\!-\!\lambda_{-n})^3}\!-\!\frac{2}{(\mu_B'\!-\!\lambda_{-n})^3}\right)}{\mu_D(\mu_B'-\mu_D)(\mu_B'+\mu_D-\lambda-\lambda_{-n})}\!+\!\frac{2\lambda_n\lambda_{-n}(\mu_D+\mu_B'-\lambda)}{\mu_B'\mu_D(\mu_D\!+\!\mu_B'\!-\!\lambda\!-\!\lambda_{-n})^3}.\label{eq:EqBjD}\\
&\Phi_{q,L_{j,D}}=\frac{\lambda_{-n}}{\mu_D(\mu_D\!-\!\lambda_{-n})}\!\left(\!\frac{1}{\lambda_n}-\frac{\lambda_n(\mu_B'+\mu_D-\lambda)}{\mu_B'(\mu_B'\!+\!\mu_D\!-\!\lambda\!-\!\lambda_{-n})^2}-\!\frac{\lambda_n(\mu_B'\!-\!\lambda)\left(\frac{1}{\mu_B'\!-\!\lambda_{-n}}-\frac{\mu_B'\!-\!\lambda_{-n}}{(\mu_D\!-\!\lambda_{-n})^2}\right)}{(\mu_D-\mu_B')(\mu_B'+\mu_D-\lambda-\lambda_{-n})}\right).\label{eq:EqLjD}\\
&\Phi_{q\!,B_{j\!,U}}\!=\!
    \frac{\lambda_n(\mu_B'-\lambda)(\mu_B'-\lambda_{-n})(\mu_D+\mu_n'-\lambda_n)}{\mu_D(\mu_D\!-\!\lambda\!+\!\mu_n')^2(\mu_n'-\lambda_n)(\mu_B'-\mu_D)(\mu_B'+\mu_D-\lambda-\lambda_{-n})}\notag\\
    &\qquad\qquad-\frac{\lambda_n\mu_D(\mu_B'-\lambda)(\mu_B'-\lambda_{-n})(\mu_D+\mu_n'-\lambda_n)}{(\mu_n'-\lambda_n)(\mu_B'-\mu_D)(\mu_B'+\mu_D-\lambda-\lambda_{-n})((\mu_D\!+\!\mu_n')\mu_B'\!-\!\lambda\mu_D\!+\!(\mu_D\!-\!\mu_B')\!\lambda_n)^2}\notag\\
    &+\frac{\lambda_n\mu_B'\mu_D\frac{(\mu_D+\mu_n'-\lambda_n)(\mu_B'+\mu_D-\lambda)}{\mu_n'-\lambda_n}}{(\!(\mu_B'\!+\!\mu_D\!-\!\lambda)(\mu_D\!+\!\mu_n'\!-\!\lambda_n)(\mu_B'\!-\!\lambda_{\!-\!n})\!+\!\lambda_{\!-\!n}(\!\mu_D^2\!+\!(2\mu_n'\!-\!2\lambda_n\!-\!\lambda)\mu_D\!+\!(\mu_B'\!-\!2\lambda)\!(\mu_n'\!-\!\lambda_n)\!)\!)^2}.\label{eq:EqBjU}\\
&\Phi_{q,L_{j,U}}=\frac{\lambda_n(\mu_D-\lambda)(\mu_D-\lambda_{-n})}{\mu_B'(\mu_n'\!-\!\lambda_n)\!(\mu_D\!-\!\mu_n'\!-\!\lambda_{-n})\!(\mu_D\!+\!\mu_n'\!-\!\lambda)}\!\left(\!\frac{\mu_B'+\mu_n'-\lambda_n}{(\mu_B'+\mu_n'-\lambda)^2}-\frac{\mu_B'+\mu_D-\lambda}{(\mu_B'\!+\!\mu_D\!-\!\lambda\!-\!\lambda_{-n})^2}\!\right)\notag\\
    &\qquad\qquad+\frac{\lambda_n(\mu_B'\!-\!\lambda)(\mu_B'\!-\!\lambda_{-n})(\mu_D\!-\!\lambda)(\mu_D\!-\!\lambda_{-n})\left(\frac{1}{\mu_n'^2}-\frac{1}{(\mu_B'-\lambda_{-n})^2}\right)}{(\mu_n'\!-\!\lambda_n)(\mu_D\!-\!\mu_n'\!-\!\lambda_{-n})(\mu_D+\mu_n'-\lambda)(\mu_B'+\mu_n'-\lambda)(\mu_B'-\mu_n'-\lambda_{-n})}\notag\\
    &\qquad\qquad-\frac{\lambda_n(\mu_B'\!-\!\lambda)(\mu_B'\!-\!\lambda_{-n})(\mu_D\!-\!\lambda)(\mu_D\!-\!\lambda_{-n})\left(\frac{1}{(\mu_D-\lambda_{-n})^2}-\frac{1}{(\mu_B'-\lambda_{-n})^2}\right)}{(\mu_n'\!-\!\lambda_n)(\mu_D\!-\!\mu_n'\!-\!\lambda_{-n})(\mu_B'-\mu_D)(\mu_B'+\mu_D-\lambda-\lambda_{-n})}.\label{eq:EqLjU}
\end{align}
\hrulefill\vspace{-2.5em}
\end{figure*}

\begin{Theorem}\label{Theorem3}
In the partial computing scheme with the offloading ratio $p$, the closed-form expression for the average AoI of $U_n$ is derived as
\begin{align}\label{eq:expreAoIpar}
    \Delta_{n,p} =& \frac{1}{\lambda_n}\!+\!\frac{1}{\mu_B'}\!+\!\frac{1}{\mu_D}\!+\!\frac{1}{\mu_n'}\!+\!\frac{\lambda_{-n}}{\mu_B'(\mu_B'\!-\!\lambda_{-n})}+\!\frac{\lambda_n^2\lambda_{-n}}{\mu_B'(\mu_B'\!-\!\lambda_{-n})^3}\notag\\
    &\!+\!\frac{\lambda_n^2}{(\mu_B'\!-\!\lambda)(\mu_B'\!-\!\lambda_{-n})^2}+\lambda_n(\Phi_{p,B_{j,D}}\!+\!\Phi_{p,L_{j,D}}\!+\!\Phi_{p,B_{j,U}}\!+\!\Phi_{p,L_{j,U}}),
\end{align}
and the closed-form expression for the average PAoI of $U_n$ is derived as
\begin{align}\label{eq:exprePAoIpar}
\Omega_{n,p} = \frac{1}{\lambda_n}+\frac{1}{\mu_B'-\lambda}+\frac{1}{\mu_D-\lambda}+\frac{1}{\mu_n'-\lambda_n},
\end{align}
where $\lambda=\sum\limits_{n=1}^N\lambda_n$ and $\lambda_{-n}=\lambda-\lambda_n$.  
Here, $\Phi_{p,B_{j,D}}$, $\Phi_{p,L_{j,D}}$, $\Phi_{p,B_{j,U}}$, and $\Phi_{p,L_{j,U}}$ are given by \eqref{eq:EqBjD}, \eqref{eq:EqLjD}, \eqref{eq:EqBjU}, and \eqref{eq:EqLjU}, respectively.

\begin{IEEEproof}
See Appendix \ref{Appendix:A}.
\end{IEEEproof}

\end{Theorem}

From Theorem \ref{Theorem3}, we find that the  average PAoI has a concise expression than the average AoI. Based on \eqref{eq:expreAoIpar}, we obtain the average AoI and the average PAoI in the local computing and the edge computing scheme by setting $p=0$ and $p=1$, which are given in the following two corollaries, respectively. 

\begin{Corollary}\label{Theorem1}
In the local computing scheme, the closed-form expression for the average AoI of $U_n$ is derived as
\begin{align}\label{eq:AoIepres}
\Delta_{n,l} \!=&\frac{1}{\lambda_n}\!+\!\frac{1}{\mu_D}\!+\!\frac{1}{\mu_n}\!+\!\frac{\lambda_{-n}}{\mu_D(\mu_D\!-\!\lambda_{-n})}+\!\frac{\lambda_n^2\lambda_{-n}}{\mu_D(\mu_D\!-\!\lambda_{-n})^3}+\frac{\lambda_n^2}{(\mu_D-\lambda)(\mu_D-\lambda_{-n})^2}\notag\\
    &+\frac{\lambda_n^2(\mu_D+\mu_n-\lambda_n)}{\mu_D(\mu_n-\lambda_n)(\mu_D+\mu_n-\lambda)^2}+\frac{\lambda_n^2(\mu_D-\lambda)(\mu_D+\mu_n-\lambda_{-n})}{\mu_n^2(\mu_D-\lambda_{-n})(\mu_n-\lambda_n)(\mu_D+\mu_n-\lambda)},
\end{align}
and the closed-form expression for the average PAoI of $U_n$ is derived as
\begin{align}\label{eq:PAoIepres}
\Omega_{n,l} = \frac{1}{\lambda_n}+\frac{1}{\mu_D-\lambda}+\frac{1}{\mu_n-\lambda_n}.
\end{align}

\begin{IEEEproof}
In this scheme, all the tasks in each packet of $U_n$ are computed by the local server at $U_n$. It can be regarded as a partial computing scheme with $p=0$, where the computation rate of the local server at $U_n$ is given as $\mu_n'=\mu_n$ and the computation rate of the edge server is given as $\mu_B'\rightarrow\infty$. Therefore, we obtain the closed-form expression for the average AoI of $U_n$ in the local computing scheme, $\Delta_{n,l}$, and the average PAoI of $U_n$ in the local computing scheme, $\Omega_{n,l}$, as
\begin{align}
    \Delta_{n,l} = \Delta_{n,p}\big|_{\mu_B'\rightarrow\infty,\mu_n'=\mu_n},
\end{align}
and
\begin{align}
    \Omega_{n,l} = \Omega_{n,p}\big|_{\mu_B'\rightarrow\infty,\mu_n'=\mu_n},
\end{align}
respectively, which are given in \eqref{eq:AoIepres} and \eqref{eq:PAoIepres}.
\end{IEEEproof}

\end{Corollary}


\begin{Corollary}\label{Theorem2}
In the edge computing scheme, the closed-form expression for the average AoI of $U_n$ is derived as

\begin{align}\label{eq:AoIedgeServer}
    \Delta_{n,e} =& \frac{1}{\lambda_n}+\frac{1}{\mu_B}+\frac{1}{\mu_D}+\!\frac{\lambda_{-n}}{\mu_B(\mu_B\!-\!\lambda_{-n})}\!+\!\frac{\lambda_n^2\lambda_{-n}}{\mu_B(\mu_B\!-\!\lambda_{-n})^3}\notag\\
    &+\frac{\lambda_n^2}{(\mu_B\!-\!\lambda)(\mu_B\!-\!\lambda_{-n})^2}\!+\lambda_n(\Phi_{e,B_{j,D}}+\Phi_{e,L_{j,D}}),
\end{align}
and the closed-form expression for the average PAoI of $U_n$ is derived as
\begin{align}\label{eq:PAoIedgeServer}
\Omega_{n,l} = \frac{1}{\lambda_n}+\frac{1}{\mu_B-\lambda}+\frac{1}{\mu_D-\lambda},
\end{align}
where $\Phi_{e,B_{j,D}}$ and $\Phi_{e,L_{j,D}}$ are given by \eqref{eq:EeBjD} and \eqref{eq:EeLjD}, respectively.

\begin{figure*}[t]
\normalsize
\begin{align}
&\Phi_{e,B_{j,D}}=\frac{\lambda_n(\mu_B+\mu_D-\lambda)}{\mu_B(\mu_D\!-\!\lambda)(\mu_B\!+\!\mu_D\!-\!\lambda\!-\!\lambda_{-n})^2}+\frac{\lambda_n(\mu_B-\lambda)(\mu_B-\lambda_{-n})\left(\frac{1}{(\mu_D-\lambda_{-n})^2}-\frac{1}{(\mu_B-\lambda_{-n})^2}\right)}{(\mu_B-\mu_D)(\mu_D-\lambda)(\mu_D+\mu_B-\lambda-\lambda_{-n})}\notag\\
     &+\frac{\lambda_n\lambda_{-n}(\mu_B-\lambda)(\mu_B-\lambda_{-n})\left(\frac{2}{(\mu_D\!-\!\lambda_{-n})^3}\!-\!\frac{2}{(\mu_B\!-\!\lambda_{-n})^3}\right)}{\mu_D(\mu_B-\mu_D)(\mu_B+\mu_D-\lambda-\lambda_{-n})}+\frac{2\lambda_n\lambda_{-n}(\mu_D+\mu_B-\lambda)}{\mu_B\mu_D(\mu_D\!+\!\mu_B\!-\!\lambda\!-\!\lambda_{-n})^3}\label{eq:EeBjD}.\\
&\Phi_{e,L_{j,D}}\!=\!\frac{\lambda_{-n}}{\mu_D\!(\mu_D\!-\!\lambda_{\!-\!n})}\!\left(\!\frac{1}{\lambda_n}\!-\!\frac{\lambda_n(\mu_B+\mu_D-\lambda)}{\mu_B(\!\mu_B\!+\!\mu_D\!-\!\lambda\!-\!\lambda_{\!-\!n})^2}\!-\!\frac{\lambda_n(\!\mu_B\!-\!\lambda)\left(\!\frac{1}{\mu_B\!-\!\lambda_{-n}}\!-\!\frac{\mu_B\!-\!\lambda_{-n}}{(\mu_D\!-\!\lambda_{-n})^2}\!\right)}{(\mu_D\!-\!\mu_B)\!(\mu_B\!+\!\mu_D\!-\!\lambda\!-\!\lambda_{-n})}\!\right).\label{eq:EeLjD}
\end{align}
\hrulefill\vspace{-2.5em}
\end{figure*}

\begin{IEEEproof}
In this scheme, all the tasks in each packet are computed by the edge server. It can be regarded as a partial computing scheme with $p=1$, where the computation rate of the local server at $U_n$ is given as $\mu_n'\rightarrow\infty$ and the computation rate of the edge server is given as $\mu_B'=\mu_B$. Therefore, we obtain the closed-form expression for the average AoI of $U_n$ in the edge computing scheme, $\Delta_{n,e}$, and the average PAoI of $U_n$ in the edge computing scheme, $\Omega_{n,e}$, as 
\begin{align}
    \Delta_{n,e} = \Delta_{n,p}\big|_{\mu_B'=\mu_B,\mu_n'\rightarrow\infty},
\end{align}
and
\begin{align}
    \Omega_{n,e} = \Omega_{n,p}\big|_{\mu_B'=\mu_B,\mu_n'\rightarrow\infty},
\end{align}
respectively, which are given in \eqref{eq:AoIedgeServer} and \eqref{eq:PAoIedgeServer}.
\end{IEEEproof}

\end{Corollary}

From Theorem~\ref{Theorem3}, Corollary~\ref{Theorem1}, and Corollary~\ref{Theorem2}, we find that the average AoI has a more complex closed-form expression than the average PAoI. Due to this complexity, it is difficult to derive the closed-form expression for the optimal offloading ratio to minimize the average AoI. Comparing with the average AoI, we find that the expression for the average PAoI is less complex and it is feasible to obtain the closed-form expression for the optimal offloading ratio to minimize the average PAoI. Moreover, we note that the results in Theorem~\ref{Theorem3} can be used to obtain the results for a single UE system by setting $\lambda=\lambda_n$ and $\lambda_{-n}=0$.



\section{AoI Performance Optimization}\label{Sec:Approx}

In this section, we consider an MEC system with homogeneous UEs, where all UEs share the same packet generation rate $\lambda_h$ and the same computation rate of the local server $\mu_h'=\frac{\mu_h}{1-p}$, i.e., $\lambda_n=\lambda_h$ and $\mu_n = \mu_h$, $\forall$ $n$. In this homogeneous scenario, due to the complex expression for the average AoI given in \eqref{eq:expreAoIpar}, it is hard to derive the closed-form expression for the optimal offloading ratio to minimize the average AoI.

To find the closed-form expression for the optimal offloading ratio to minimize the average AoI, we first derive the simplified upper bound and lower bound on the average AoI.

\begin{Theorem}\label{Theorem6}
In the MEC system under the aforementioned assumptions, the simple form lower bound and upper bound on the average AoI are given as
\begin{align}\label{eq:AoIlower}
    \Delta_{low} &=\Omega-\Delta_{gap},
\end{align}
and
\begin{align}
    \Delta_{up} =\Omega,
\end{align}
respectively, where $\Omega=\frac{p}{\mu_B-p\lambda}+\frac{1}{\mu_D-\lambda}+\frac{1-p}{\mu_h-(1-p)\lambda_h}$ and 
\begin{align}\label{eq:deltagap}
    \Delta_{gap} =
    &\frac{\lambda_h}{(\mu_B'-\lambda_{-h})^2}+\frac{\lambda_h}{(\mu_D-\lambda_{-h})^2}+\frac{\lambda_h}{\mu_h'^2}-\frac{\lambda_h^2\lambda_{-h}}{\mu_B'(\mu_B'-\lambda_{-h})^3}-\frac{\lambda_h^2\lambda_{-h}}{\mu_D(\mu_D-\lambda_{-h})^3}.
\end{align}
In \eqref{eq:deltagap}, $\lambda_{-h}=(N-1)\lambda_h$.
\begin{IEEEproof}
See Appendix \ref{Appendix:B}.
\end{IEEEproof}
\end{Theorem}

Here we denote $\gamma$ as the ratio between $\Delta_{gap}$ and $\Omega$, i.e., $\gamma=\frac{\Delta_{gap}}{\Omega}$. From Theorem \ref{Theorem6}, we find that $\gamma$ is negligible, i.e., $\gamma\approx0$, when the transmission rate and the computation rates of both the edge server and the local server at each UE are much larger than the packet generation rate. Under this condition, the average AoI is close to the average PAoI, i.e., $\Delta\approx\Omega$. Hence, we derive the closed-form expression for the optimal offloading ratio to minimize the average PAoI, which is approximated as the optimal offloading ratio to minimize the average AoI. 

\begin{Theorem}\label{Theorem5}
In the MEC system, an optimal offloading ratio, $p_{opt}$, to minimize the average PAoI is derived as
\begin{align}\label{eq:optp}
    p_{opt}=\left\{
    \begin{aligned}
    &0,&&\textrm{if }\mu_B\leq\frac{(\mu_h-\lambda_h)^2}{\mu_h},\\
    &1,&&\textrm{if }\mu_h\leq\frac{(\mu_B-\lambda)^2}{\mu_B},\\
    &\frac{\sqrt{\mu_B\mu_h}+\lambda_h-\mu_h}{\left(1+N\sqrt{\frac{\mu_h}{\mu_B}}\right)\lambda_h},&&\textrm{otherwise}.
    \end{aligned}
    \right.
\end{align}
\begin{IEEEproof}
See Appendix \ref{Appendix:C}.
\end{IEEEproof}
\end{Theorem}

From Theorem \ref{Theorem5}, we see that the optimal offloading ratio increases as the computation rate of the edge server increases, but decreases as the computation rate of the local server increases. 

\section{Numerical Results}\label{sec:Numerical}
In this section, we present numerical results to validate our analysis in Section~\ref{sec:Derivation} and ~\ref{Sec:Approx}. In particular, we present numerical results in the homogeneous case where all UEs share the same packet generation rate $\lambda_h$ and the same local computation rate $\mu_h$, i.e., $\lambda_n=\lambda_h$ and $\mu_n=\mu_h$, for $\forall n$. 

\begin{figure}[!t]
\centering
\includegraphics[width=0.75\columnwidth]{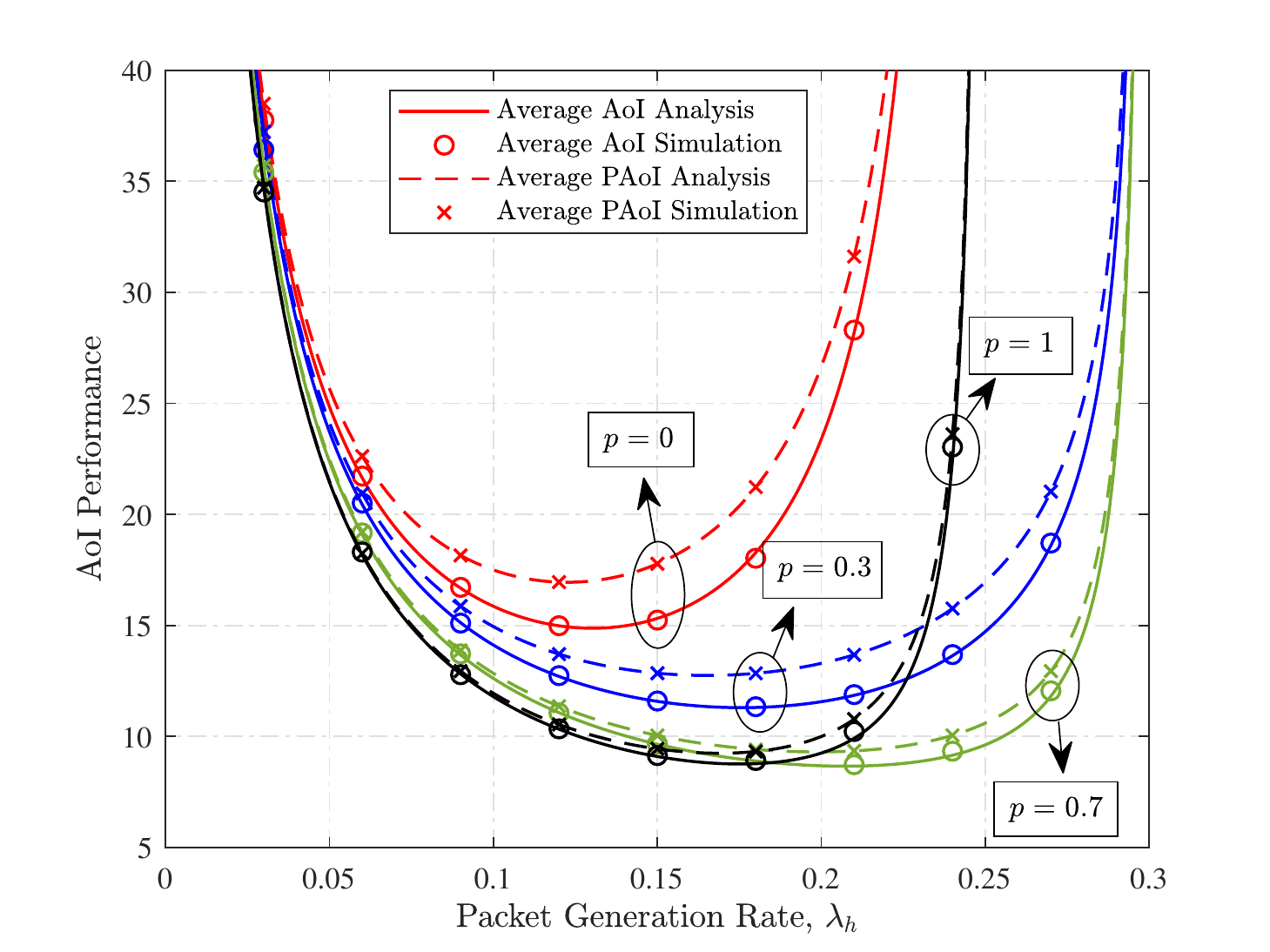}
\vspace{-2em}
\caption{The average AoI and PAoI of the MEC system versus the packet generation rate, $\lambda_h$, with $N=6$, $\mu_B=1.5$, $\mu_h=0.25$, and $\mu_D=1.8$.}\label{fig:2ps}
\vspace{-2.5em}
\end{figure}

Fig.~\ref{fig:2ps} plots the average AoI and PAoI of the MEC system versus the packet generation rate, $\lambda_h$. We first observe that the analytical average AoI and PAoI of the MEC system tightly match the simulation results, which demonstrates the correctness of our analytical result. We then observe that for all considered values of $p$, the average AoI and PAoI of the MEC system first decrease and then increases when $\lambda_h$ increases, where the average AoI has a same trend with the average PAoI vesus the packet generation rate, $\lambda_h$. 
This observation is due to the fact that the increase in $\lambda_h$ has a two-fold effect on the average AoI and PAoI of the MEC system. When $\lambda_h$ is small, this increase leads to a higher updating rate of packets, which decreases the average AoI and PAoI of the MEC system. 
When $\lambda_h$ exceeds a certain threshold, its increase leads to the significant increase in the waiting time of a packet in computation queues and the transmission queue, thereby degrading the AoI performance. We further observe that for small $\lambda_h$, the average AoI and PAoI of the MEC system monotonically decreases with increasing $p$. However, for large $\lambda_h$, the average AoI and PAoI of the MEC system firstly decreases and then increases with increasing $p$. This is because that when $\lambda_h$ is small, the increase in $p$ leads to a higher computation rate of the edge server, which reduces the average AoI and PAoI of the MEC system. When $\lambda_h$ is very large, the increase in $p$ results in a significantly long waiting time of a packet in the computation queue at the edge server, which increases the average AoI and PAoI of the MEC system. In addition, we find that the average AoI and PAoI is small in the partial computing scheme, but large in the other two schemes for large $\lambda_h$. This observation implies that by carefully partitioning the computational tasks, the partial computing scheme has a higher system stability than both the local and the edge computing schemes.

\begin{figure}[!t]
\centering
\includegraphics[width=0.75\columnwidth]{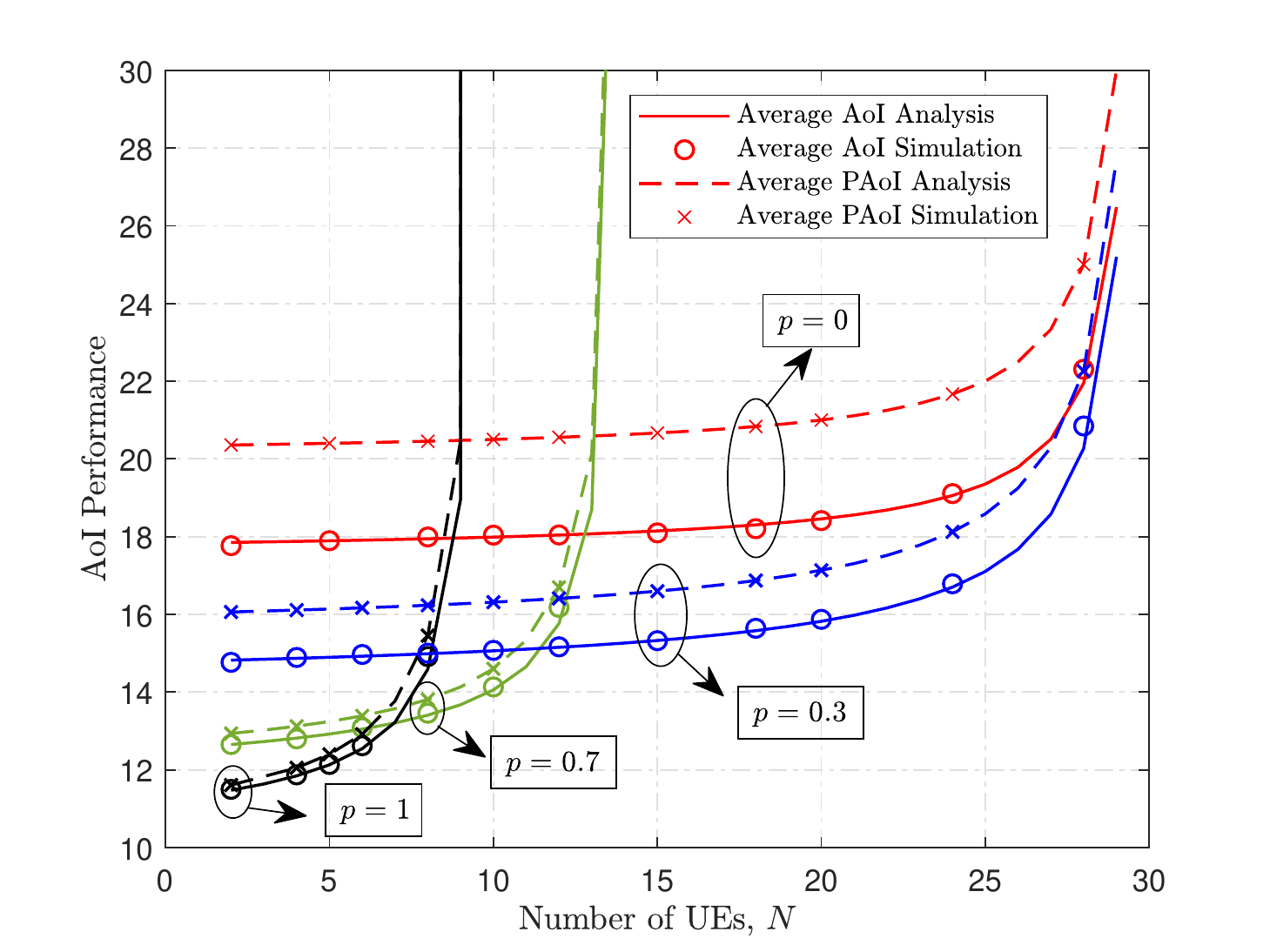}
\vspace{-2em}
\caption{The average AoI and PAoI of the MEC system versus the number of UEs, $N$, with $\lambda_h=0.1$, $\mu_B=1$, $\mu_h=0.2$, and $\mu_D=3$.}\label{fig:3ps}
\vspace{-2.5em}
\end{figure}

Fig.~\ref{fig:3ps} plots the average AoI and PAoI of the MEC system versus the number of UEs, $N$. We first observe that when $N$ increases, the average AoI and PAoI of the MEC system increase monotonically and this increase is faster for larger $p$. This is because that the increase in $N$ results in the longer waiting time of a packet in both the transmission queue and the computation queue at the edge server, which increases the average AoI and PAoI of the MEC system. If $p$ is large, the tasks computed by the edge server increases dramatically as $N$ increases, which results in a long waiting time of a packet in the computation queue. We further observe that the edge computing scheme has a lower average AoI and PAoI than the local computing scheme and the partial computing scheme for small $N$, but a larger average AoI and PAoI for large $N$. 
This is because that for a small number of UEs, compared to local computing, edge computing can provide the higher computation rate via the powerful edge server, thereby decreasing the average AoI and PAoI of the MEC system. Differently, for a large number of UEs, allocating most computational tasks to the local server can avoid the long waiting time in the computation queue at the edge server, which is beneficial to the decrease in the average AoI and PAoI. 
This observation also indicates that the careful design of the offloading ratio in the partial computing scheme can effectively decrease the average AoI and PAoI, especially when the number of UEs is large. 


\begin{figure}[!t]
\centering
\includegraphics[width=0.75\columnwidth]{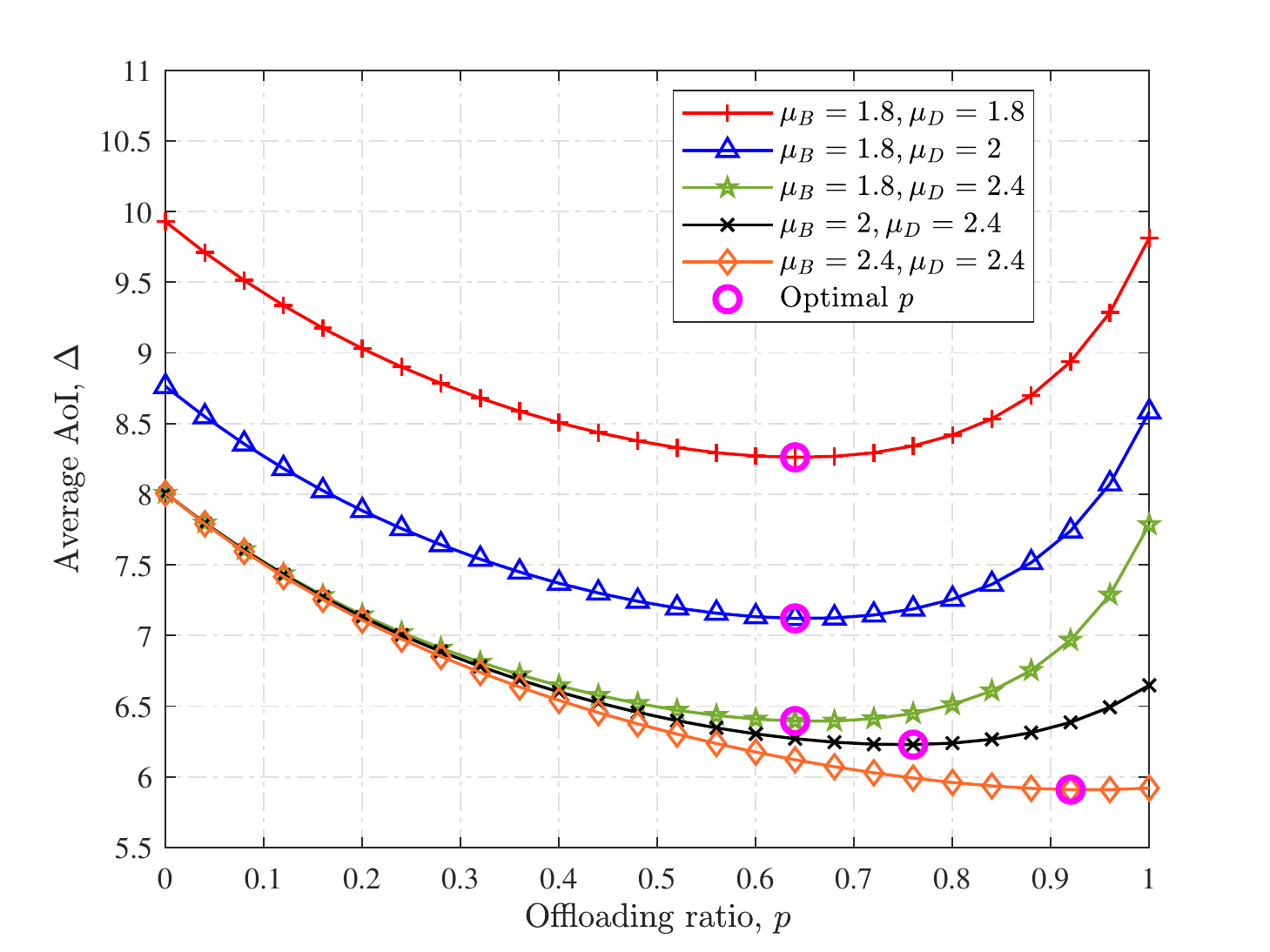}
\vspace{-2em}
\caption{The average AoI of the MEC system versus the offloading ratio, $p$, with $N=6$, $\lambda_h=0.25$, and $\mu_h=0.5$.}\label{fig:4ps}
\vspace{-2.5em}
\end{figure}
                       
Fig.~\ref{fig:4ps} plots the average AoI of the MEC system versus the offloading ratio, $p$, for different $\mu_D$ and $\mu_B$. We first observe that the average AoI first decreases and then increases as $p$ increases. 
To understand the trend observed in these curves, we note that the increase in $p$ leads to the decrease in the waiting time in the computation queue at the local server but the increase in the waiting time in the computation queue at the edge server. When $p$ is small, its increase significantly reduces the waiting time of a packet at the local server, which decreases the average AoI of the MEC system. When the value of $p$ increases up to a certain threshold, the long waiting time of a packet at the edge server dominates the average AoI, thereby resulting in the increase in the average AoI. We then observe that when $\mu_D$ increases, the average AoI of the MEC system decreases, while the optimal $p$ almost remains the same value. This is because that as $\mu_D$ increases, the high transmission rate of the BS leads to reduced waiting time in the transmission queue, which decreases the average AoI. However, the high transmission rate of the BS only affects the queuing delay of a packet in the transmission queue, while the optimal $p$ 
is independent of the transmission queue. 
We further observe that when $\mu_B$ increases, the average AoI of the MEC system decreases and the optimal $p$ increases. This is because that the larger value of $\mu_B$ means the higher computation rate of the edge server, which can reduce the waiting time in the computation queue at the edge server and then lead to decreased average AoI. In addition, when the edge server has a higher computation rate, it can decrease the average AoI by allocating more computational tasks to the edge server.



\begin{figure}[!t]
\centering
\includegraphics[width=0.75\columnwidth]{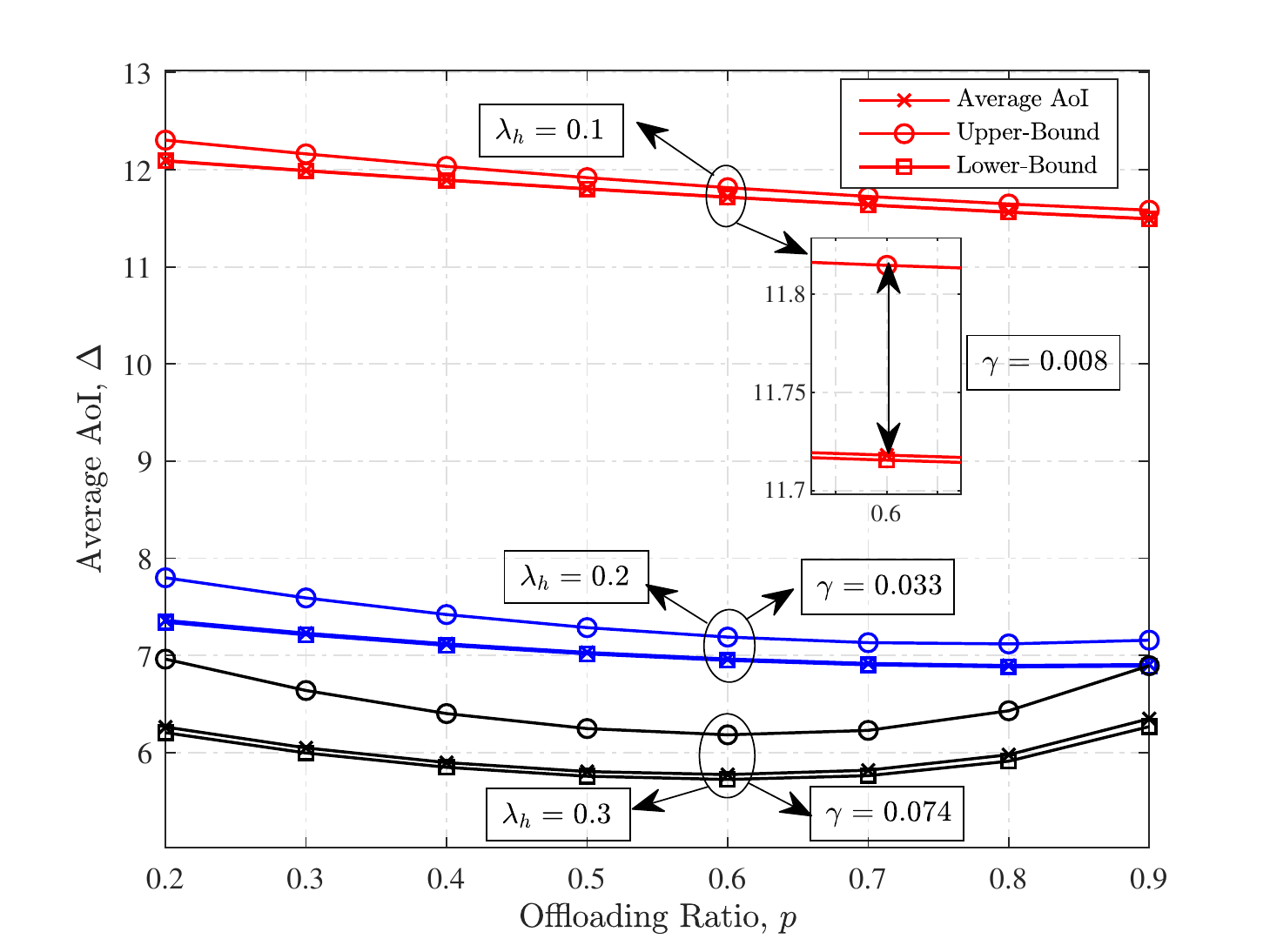}
\vspace{-2em}
\caption{The average AoI of the MEC system versus the offloading ratio, $p$, with $N=4$, $\mu_B=1.5$, $\mu_h=0.6$, and $\mu_D=2$.}\label{fig:6ps}
\vspace{-2.5em}
\end{figure}

Fig.~\ref{fig:6ps} plots the upper bound and the lower bound on the average AoI in Theorem \ref{Theorem6}. 
We first observe that the average AoI is lower then the upper bound and higher then the lower bound given in Theorem \ref{Theorem6}, which demonstrates the correctness of our analytical result. We then observe that $\gamma$ is small when $\lambda_h$ is small and it is large for a large $\lambda_h$. This observation is due to the fact that the decrease in $\lambda_h$ leads to the decrease in the gap between the upper bound and the lower bound, which decreases $\gamma$. This observation implies that the expression for the average PAoI can be adopted as a reasonable approximation of the average AoI when the serving rate is much greater than the packet generation rate.


\begin{figure}[!t]
\centering
\includegraphics[width=0.75\columnwidth]{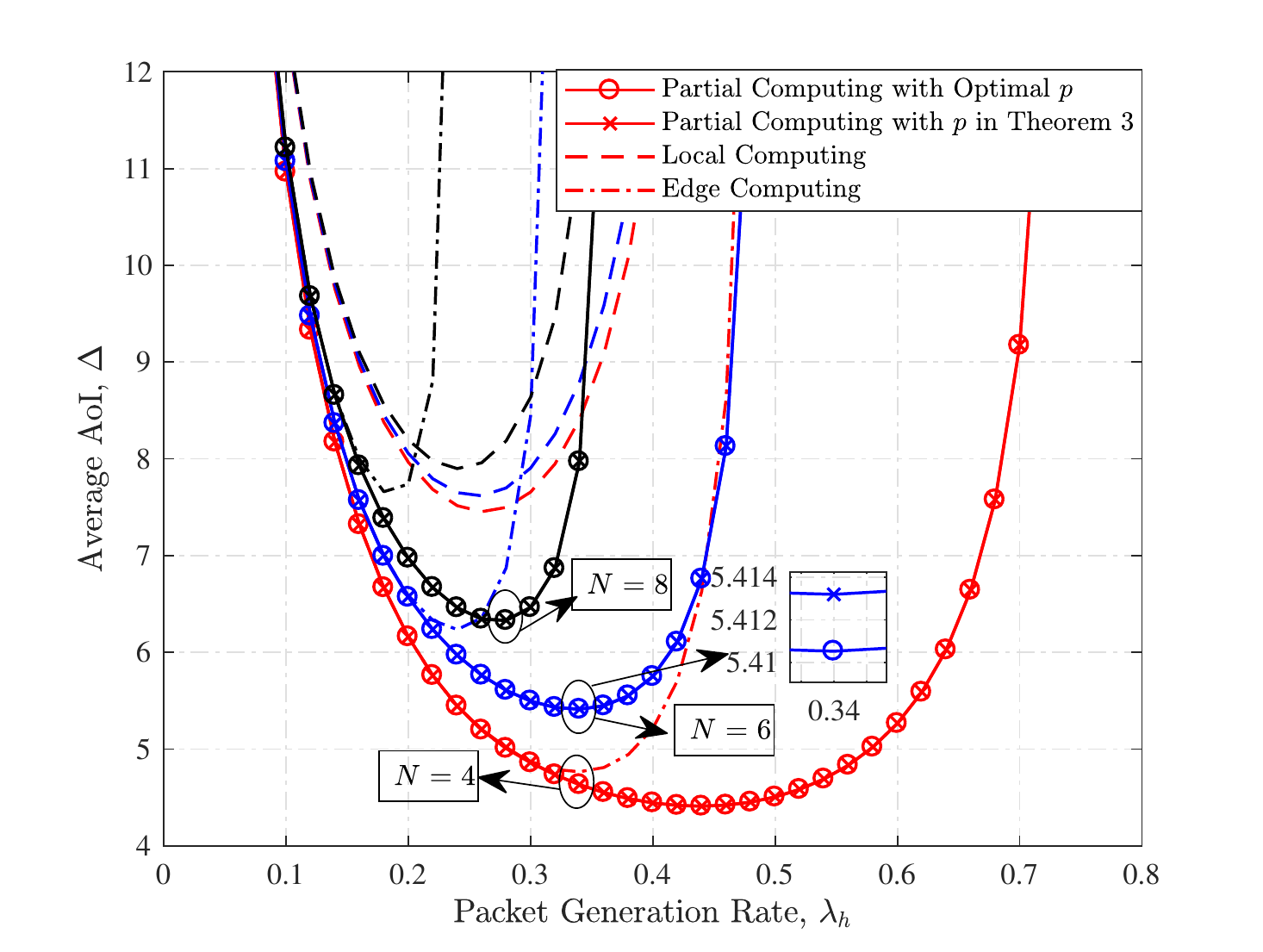}
\vspace{-2em}
\caption{The average AoI of the MEC system versus the packet generation rate, $\lambda_h$, with $\mu_B=2$, $\mu_h=0.5$, and $\mu_D=3$.}\label{fig:7ps}
\vspace{-2.5em}
\end{figure}

Fig.~\ref{fig:7ps} plots the average AoI achieved by the partial computing scheme with the optimal offloading ratio, the partial computing scheme with the offloading ratio in Theorem~\ref{Theorem5}, the local computing scheme, and the edge computing scheme of the MEC system versus the packet generation rate, $\lambda_h$, for different $N$. We observe that the considered partial computing scheme with the offloading ratio in Theorem~\ref{Theorem5} achieves the lowest average AoI comparing with the local computing scheme and the edge computing scheme, especially for large $\lambda_h$. Additionally, we observe that the gap between the optimal average AoI and the average AoI with the offloading ratio in Theorem~\ref{Theorem5} is negligible no matter $\lambda_h$ is small or large. It implies that the optimal offloading ratio, $p_{opt}$, to minimize the average PAoI can be adopted to minimize the average AoI of an MEC system, regardless of the packet generation rate.

\section{Conclusion}\label{sec:Conclusion}

We analyzed the AoI performance of a multi-user MEC system. In this MEC system, we considered three computing schemes, i.e., the local computing scheme, the edge computing scheme, and the partial computing scheme. For the three computing schemes, we derived the closed-form expressions for the average AoI, where the packets are served in both the computation queue and the transmission queue according to the FCFS policy. In the MEC system with a sufficiently large number of UEs, we derived the closed-form expression for the average AoI. Due to the complex expression for the average AoI, we further provided a simple expression upper bound and lower bound on the average AoI. Considering the sporadic packet generation pattern, we found that the gap between the upper bound and the lower bound is negligible, such that the average AoI can be approximated by the average PAoI. In addition, we derived the optimal offloading ratio to minimize the average PAoI and found that this offloading ratio can be adopted to minimize the average AoI.

\begin{appendices}

\section{Proof for Theorem \ref{Theorem3}}\label{Appendix:A}
The proof of Theorem \ref{Theorem3} relies on the following properties of exponential random variables, Poisson processes, and the $M/M/1$ queue.
\begin{Lemma}\label{Lemma1}
Let us consider that $X_1$ and $X_2$ are independent exponential random variables with $\E[X_i] = 1/\gamma_i$. Given $X_1<X_2$, the conditional PDF of variable $V=X_2-X_1$ is given by 
\begin{align}
   f_{V|X_1<X_2}(v|X_1<X_2) = \gamma_2 \exp(-\gamma_2 v),\ v\geq 0. 
\end{align}
\end{Lemma}

\begin{Lemma}\label{Lemma2}
Given a Poisson process $K(t)$ with the rate $\lambda$ and an exponential random variable $X$ with $\E[X]=1/\gamma$, the number of arrivals $K(X)$ in the interval $[0,X]$ has the geometric probability mass function (PMF), which is given by
\begin{align}
    \mathrm{Pr}(K(X)=k) = (1-\alpha)\alpha^k,\ k\geq 0,
\end{align}
where $\alpha=\lambda/(\gamma+\lambda)$.
\end{Lemma}

We first derive the average AoI of $U_n$. Based on \eqref{eq:newAoIevox}, the average AoI of $U_n$ in the partial computing scheme is calculated as
\begin{align}\label{eq:Deltanp}
     \Delta_{n,p}
     =&\frac{1}{\lambda_n}+\frac{1}{\mu_B'}+\frac{1}{\mu_D}+\frac{1}{\mu_n'}+\!\lambda_n\left(\E[Y_jW_{j,B}]+\E[Y_jW_{j,D}]+\E[Y_jW_{j,U}]\right).
\end{align}

To obtain $\Delta_{n,p}$, we need to derive $\E[Y_jW_{j,B}]$, $\E[Y_jW_{j,D}]$, and $\E[Y_jW_{j,U}]$ in \eqref{eq:Deltanp}. We first derive $\E[Y_jW_{j,B}]$ as
\begin{align}\label{E1YjWj1}
    \E[Y_jW_{j,B}]=&\E[Y_jW_{j,B}|B_{j,B}]\mathrm{Pr}(B_{j,B})+\E[Y_jW_{j,B}|L_{j,B}]\mathrm{Pr}(L_{j,B}),
\end{align}
where $B_{j,B}$ denotes the event that $P_{n,j}$ is generated before $P_{n,j-1}$ arrives at the transmission queue and $L_{j,B}$ denotes the event that $P_{n,j}$ is generated after $P_{n,j-1}$ arrives at the transmission queue. Here, $B_{j,B}$ and $L_{j,B}$ are two complementary events such that $\mathrm{Pr}(B_{j,B})+\mathrm{Pr}(L_{j,B})=1$. We first calculate $\E[Y_jW_{j,B}|B_{j,B}]\mathrm{Pr}(B_{j,B})$ in \eqref{E1YjWj1}. When $B_{j,B}$ happens, $W_{j,B}$ is calculated as
\begin{align}\label{eq:WjB}
    W_{j,B} = X_{j-1,B}-Y_{j}+\sum\limits_{\kappa=1}^{K_{j,y}} S_{j,\kappa,B},
\end{align}
where $K_{j,y}$ is the number of packets generated for other UEs during $Y_j$ and $S_{j,\kappa,B}$ is the computation time of the $\kappa$th packet among $K_{j,y}$ packets computed by the edge server. As $X_{j-1,B}$ and $Y_j$ are independent exponential random variables, according to Lemma \ref{Lemma1}, we obtain
\begin{align}\label{eq:fW1BjB}
    f_{V_j|B_{j,B}}(v|B_{j,B})=(\mu_B'-\lambda)\exp(-(\mu_B'-\lambda)v),
\end{align}
where $V_j=X_{j-1,B}-Y_j$. In addition, as the packet of each UE is generated according to a Poisson process, we obtain the conditional PMF of the number of packets generated during $Y_j=y$ as
\begin{align}\label{eq:Ky}
    \mathrm{Pr}(K_{j,y}=k|Y_j=y) = \frac{(\lambda_{-n}y)^k\exp(-\lambda_{-n}y)}{k!}.
\end{align}
Combining \eqref{eq:WjB}, \eqref{eq:fW1BjB}, with \eqref{eq:Ky}, we obtain
\begin{align}\label{eq:WjBgivenYjBjB}
    \E[W_{j,B}|Y_j=y,B_{j,B}] = \frac{1}{\mu_B'-\lambda}+\frac{\lambda_{-n}y}{\mu_B'}.
\end{align}
According to \eqref{eq:WjBgivenYjBjB}, we calculate $\E[Y_jW_{j,B}|B_{j,B}]\mathrm{Pr}(B_{j,B})$ as
\begin{align}\label{eq:EBJ1case1}
    \E[Y_jW_{j,B}|B_{j,B}]\mathrm{Pr}(B_{j,B}) 
    &=\int_{0}^{\infty}yf_{Y_j}(y)\mathrm{Pr}(B_{j,B}|Y_j\!=\!y)\E[W_{j,B}|Y_j=y,B_{j,B}] \mathrm{d}y\notag\\
    &= \frac{\lambda_n}{(\mu_B'-\lambda_{-n})^2}\left(\frac{2\lambda_{-n}}{\mu_B'(\mu_B'-\lambda_{-n})}+\frac{1}{\mu_B'-\lambda}\right).
\end{align}
We then calculate $\E[Y_jW_{j,B}|L_{j,B}]\mathrm{Pr}(L_{j,B})$. When $L_{j,B}$ happens, $W_{j,B}$ is calculated as
\begin{align}\label{eq:WLjB}
    W_{j,B} = \sum\limits_{\kappa=1}^{K_{j,e}} S_{j,\kappa,B},
\end{align}
where $K_{j,e}$ is the number of packets in the computation queue at the edge server when $P_{n,j}$ is generated. Based on Lemma \ref{Lemma2}, we obtain that when $P_{n,j}$ is generated, the PMF of the number of packets in the computation queue at the edge server is given by
\begin{align}\label{eq:PKje}
    \mathrm{Pr}(K_{j,e}=k) = \left(\frac{\lambda_{-n}}{\mu_B'}\right)^k\left(1\!-\!\frac{\lambda_{-n}}{\mu_B'}\right).
\end{align}
Combining \eqref{eq:WLjB} with \eqref{eq:PKje}, we obtain
\begin{align}\label{eq:WjBgivenYjLjB}
    \E[W_{j,B}|Y_j=y,L_{j,B}] = \frac{\E[K_{j,e}]}{\mu_B'} = \frac{\lambda_{-n}}{\mu_B'(\mu_B'-\lambda_{-n})}.
\end{align}
According to \eqref{eq:WjBgivenYjLjB}, we calculate the second item in \eqref{E1YjWj1} as  
\begin{align}\label{eq:ELJ1case1}
    \E[Y_jW_{j,B}|L_{j,B}]\mathrm{Pr}(\!L_{j,B}\!)
    &=\int_{0}^{\infty}yf_{Y_j}(y)\mathrm{Pr}(L_{j,B}|Y_j\!=\!y)\E[W_{j,B}|Y_j=y,L_{j,B}] \mathrm{d}y\notag\\
    &=\frac{\lambda_{-n}}{\mu_B'(\mu_B'-\lambda_{-n})}\left(\frac{1}{\lambda_n}-\frac{\lambda_n}{(\mu_B'-\lambda_{-n})^2}\right).    
\end{align}
By substituting \eqref{eq:EBJ1case1} and \eqref{eq:ELJ1case1} into \eqref{E1YjWj1}, we obtain $\E[Y_jW_{j,B}]$.    

Next, we derive $\E[Y_jW_{j,D}]$ in \eqref{eq:Deltanp}. We denote $Y_{j,B}$ as the time interval when $P_{n,j-1}$ and $P_{n,j}$ arrive at the transmission queue, i.e., $Y_{j,B} = t_{j,B}-t_{j-1,B}$. We then denote $B_{j,D}$ as the event that $P_{n,j}$ arrives at the transmission queue before $P_{n,j-1}$ arrives at the computation queue at the local server at $U_n$ and $L_{j,D}$ as the event that $P_{n,j}$ arrives at the transmission queue after $P_{n,j-1}$ arrives at the computation queue at the local server at $U_n$. Based on these two complementary events, we calculate $\E[Y_jW_{j,D}]$ as  
\begin{align}\label{eq:sub2YW2}
    \E[Y_jW_{j,D}]\!=\Phi_{p,B_{j,D}}+\Phi_{p,L_{j,D}},
\end{align}
where $\Phi_{p,B_{j,D}}=\E[Y_jW_{j,D}|{B_{j,D}}]\mathrm{Pr}({B_{j,D}})$ and $\Phi_{p,L_{j,D}}=\E[Y_jW_{j,D}|{L_{j,D}}]\mathrm{Pr}({L_{j,D}})$.

Here, we first calculate $\Phi_{p,B_{j,D}}$. When $B_{j,D}$ happens, $W_{j,D}$ depends on both the residual queuing delay of $P_{n,j-1}$ in the transmission queue and the transmission time of the packets generated during $Y_j$. Thus, we calculate $\Phi_{p,B_{j,D}}$ as
\begin{align}\label{eq:E1YWj2Bj2}
   \Phi_{p,B_{j,D}}=&\E[Y_j(X_{j\!-\!1,D}\!-\!Y_{j,B})|B_{j,D}]\mathrm{Pr}(B_{j,D}) +\!\E\left[Y_j\sum\limits_{\kappa=1}^{K_{j,y}} S_{j,\kappa,D}\bigg|B_{j,D}\right]\mathrm{Pr}(B_{j,D}),
\end{align}
where $S_{j,\kappa,D}$ denotes the transmission time of the $\kappa$th packet among $K_{j,y}$ packets. Based on Lemma \ref{Lemma1}, we obtain
\begin{align}\label{eq:EXj1DYjBjD}
    \E[X_{j\!-\!1,D}\!-\!Y_{j,B}|B_{j,D}] = \frac{1}{\mu_D-\lambda}.
\end{align}
Since $\mathrm{Pr}(B_{j,D})$ is calculated as
\begin{align}\label{eq:PrBjD}
   &\mathrm{Pr}(B_{j,D}) \!=\! \int_0^{\infty}\!\int_0^{\infty}f_{Y_j\!,Y_{j\!,B}}(y\!,y')\exp(-\!(\mu_D\!-\!\lambda)y')\mathrm{d}y'\mathrm{d}y,
\end{align}
we need to derive the conditional PDF $f_{Y_{j,B}|Y_j}(y'|y)$ to obtain $\mathrm{Pr}(B_{j,D})$.   
Here, $Y_{j,B}$ is calculated as
\begin{align}
    Y_{j,B} = X_{j,B}+Y_j-X_{j-1,B},
\end{align}
where the PDF of $X_{j\!-\!1,B}$ is given as
\begin{align}\label{eq:fXj1}
    f_{X_{j\!-\!1,B}}(x_1) = (\mu_B'-\lambda)\exp(-(\mu_B'-\lambda)x_1).
\end{align}
We consider two complementary events, $B_{j,B}$ and $L_{j,B}$, and derive $f_{Y_{j,B}|Y_j}(y'|y)$ as
\begin{align}\label{eq:fYjBgivenYj}
     f_{Y_{j,B}|Y_j}(y'|y)=p_{Y_{j,B},B_{j,B}|Y_j}+p_{Y_{j,B},L_{j,B}|Y_j},
\end{align}
where $p_{Y_{j,B},B_{j,B}|Y_j}$ and $p_{Y_{j,B},L_{j,B}|Y_j}$ are given by
\begin{align}
    &p_{Y_{j,B},B_{j,B}|Y_j} =  f_{Y_{j,B}|Y_j,B_{j,B}}(y'|y,B_{j,B})\mathrm{Pr}(B_{j,B})
\end{align}
and  
\begin{align}
    p_{Y_{j,B},L_{j,B}|Y_j} =  f_{Y_{j,B}|Y_j,L_{j,B}}(y'|y,L_{j,B})\mathrm{Pr}(L_{j,B}),
\end{align}
respectively. 


When $B_{j,B}$ happens, $Y_{j,B}$ depends on the computation time of the packets generated during $Y_j$ at the edge server. We consider that there are $K_{j,y}=k$ packets generated during $Y_j$, where the PMF of $k$ packets generated during $Y_j$ is given by \eqref{eq:Ky}. As the computation time of each packet follows the independent and identical exponential distribution, the total time consumed to compute these $k$ packets and $P_{n,j}$ follows a Gamma distribution, whose PDF is given by
\begin{align}\label{eq:fYjD}
    f_{Y_{j,B}|K_{j,y}}(y'|k) = \frac{y'^k\mu_B'^{k+1}\exp(-\mu_B' y')}{k!}.
\end{align}
Combining \eqref{eq:Ky}, \eqref{eq:fXj1}, with \eqref{eq:fYjD}, we calculate $p_{Y_{j,B},B_{j,B}|Y_j}$ as
\begin{align}\label{eq:fy'case11}
    p_{Y_{j,B},B_{j,B}|Y_j} 
    &=\int_{y}^{\infty}f_{X_{j\!-\!1,D}}(x_1)\sum\limits_{k=0}^{\infty}\mathrm{Pr}(K_{j,y}=k|Y_j=y)f_{Y_{j,B}|K_{j,y}}(y'|k)\mathrm{d}x_1\notag\\
    &=\mu_B'\exp(-(\mu_B'-\lambda_n)y-\mu_B'y'){I_0\left(2\sqrt{\lambda_{-n}\mu_B'yy'}\right)},
\end{align}
 where $I_0(\cdot)$ is the modified first-kind Bessel function of the zeroth order.

When $L_{j,B}$ happens, $Y_{j,B}$ depends on the time interval $Y_j-X_{j-1,B}$ and $X_{j,B}$. In particular, $X_{j,B}$ depends on the number of the packets in the computation queue at the edge server when $P_{n,j}$ is generated. We consider that when $P_{n,j}$ is generated, there are $K_{j,e}=k$ packets in the computation queue at the edge server. As the computation time of each packet follows the independent and identical exponential distribution, the total time consumed to compute these $k$ packets and $P_{n,j}$ follows a Gamma distribution, whose PDF is given by
\begin{align}\label{eq:fXjB}
    f_{X_{j,B}|K_{j,e}}(x_2|k) = \frac{x_2^k\mu_B'^{k+1}\exp(-\mu_B' x_2)}{k!}.
\end{align}
Combining \eqref{eq:PKje}, \eqref{eq:fXj1}, with \eqref{eq:fXjB}, we calculate $p_{Y_{j,B},L_{j,B}|Y_j}$ as
\begin{align}\label{eq:fy'case12}
    p_{Y_{j,B},L_{j,B}|Y_j}
    &\!=\!\frac{(\mu_B'\!-\!\lambda)(\mu_B'\!-\!\lambda_{-n})}{(2\mu_B'\!-\!\lambda\!-\!\lambda_{-n})}\big(\!\exp(\!-\!(\!\mu_B'\!-\!\lambda)(y\!-\!y'))\!-\!\exp(-(\mu_B'\!-\!\lambda)y\!-\!(\mu_B'\!-\!\lambda_{-n})y')\big),
\end{align}
for $y'<y$, and 
\begin{align}\label{eq:fy'case13}
  p_{Y_{j,B},L_{j,B}|Y_j}
    &\!=\!\frac{(\mu_B'\!-\!\lambda)\!(\mu_B'\!-\!\lambda_{\!-\!n})}{(2\mu_B'\!-\!\lambda\!-\!\lambda_{\!-\!n})}\big(\!\exp(-(\!\mu_B'\!-\!\lambda_{\!-\!n\!})(y'\!-\!y))\!-\!\exp(-(\mu_B'\!-\!\lambda)y\!-\!(\mu_B'\!-\!\lambda_{-n})y')\big),
\end{align}
for $y'\geq y$. Combining \eqref{eq:fy'case11}, \eqref{eq:fy'case12}, with \eqref{eq:fy'case13}, we obtain $f_{Y_{j,B}|Y_j}(y'|y)$ in \eqref{eq:fYjBgivenYj}. By substituting \eqref{eq:fYjBgivenYj} into \eqref{eq:PrBjD} and combining \eqref{eq:PrBjD} with \eqref{eq:EXj1DYjBjD}, we obtain the first item in \eqref{eq:E1YWj2Bj2} as      
\begin{align}\label{eq:E1YWj2BJ21}
    \E[Y_j(X_{j\!-\!1,D}\!-\!Y_{j,B})|B_{j,D}]\mathrm{Pr}(B_{j,D})
    =&\frac{\lambda_n(\mu_B'+\mu_D-\lambda)}{\mu_B'(\mu_D-\lambda)(\mu_B'+\mu_D-\lambda-\lambda_{-n})^2}\notag\\
    &\!+\!\frac{\lambda_n(\mu_B'\!-\!\lambda)(\mu_B'\!-\!\lambda_{-n})\left(\frac{1}{(\mu_D\!-\!\lambda_{-n})^2}\!-\!\frac{1}{(\mu_B'\!-\!\lambda_{-n})^2}\right)}{(\mu_B'-\mu_D)(\mu_D-\lambda)(\mu_D+\mu_B'-\lambda-\lambda_{-n})}.
\end{align}
We further calculate the second item in \eqref{eq:E1YWj2Bj2}. As the PMF of $K_{j,y}$ is given by \eqref{eq:Ky}, we obtain
\begin{align}\label{eq:E1Wj21}
    \E\!\left[\!Y_j\!\sum\limits_{\kappa=1}^{K_{j\!,y}} S_{j\!,\kappa\!,D}\!\bigg|\!B_{j,D}\!\right]\!\mathrm{Pr}(B_{j,D})=&\frac{2\lambda_n\lambda_{-n}(\mu_D\!+\!\mu_B'\!-\!\lambda)}{\mu_B'\mu_D(\mu_D\!+\!\mu_B'\!-\!\lambda\!-\!\lambda_{\!-\!n\!})^3}\notag\\
    &\!+\!\frac{\lambda_n\lambda_{\!-\!n}(\mu_B'\!-\!\lambda)(\mu_B'\!-\!\lambda_{\!-\!n})\!\left(\!\frac{2}{(\mu_D\!-\!\lambda_{\!-\!n})^3}\!-\!\frac{2}{(\mu_B'\!-\!\lambda_{\!-\!n})^3}\!\right)}{\mu_D(\mu_B'-\mu_D)(\mu_B'+\mu_D-\lambda-\lambda_{-n})}.
\end{align}
By substituting \eqref{eq:E1Wj21} and \eqref{eq:E1YWj2BJ21} into \eqref{eq:E1YWj2Bj2}, we obtain $\Phi_{p,B_{j,D}}$ given by \eqref{eq:EqBjD}. 

We then calculate $\Phi_{p,L_{j,D}}$ in \eqref{eq:sub2YW2}. When $L_{j,D}$ happens, $W_{j,D}$ depends on the transmission time of the packets in the transmission queue when $P_{n,j}$ arrives at the transmission queue. Based on Lemma \ref{Lemma2}, we obtain
\begin{align}
    \E[W_{j,D}|Y_j=y,L_{j,D}] &= \E\left[\sum\limits_{\kappa=1}^{K_{j,T}} S_{j,\kappa,D}\bigg|Y_j=y,L_{j,D}\right]= \frac{\lambda_{-n}}{\mu_D(\mu_D-\lambda_{-n})},
\end{align}
where $K_{j,T}$ is the number of packets in the transmission queue when $P_{n,j}$ arrives at the transmission queue. Thus, we calculate $\Phi_{p,L_{j,D}}$ as
\begin{align}\label{eq:ELJ2case2}
    \Phi&_{p,L_{j,D}}\!=\!\int_{0}^{\infty}\!\int_{0}^{\infty}\! \E[W_{j,D}|Y_j\!=\!y,L_{j,D}]yf_{Y_j}(y) f_{Y_{j,B}|Y_j}(y'|y)\mathrm{Pr}(L_{j,D}|Y_j\!=\!y,Y_{j,B}\!=\!y')\mathrm{d}y'\mathrm{d}y,
\end{align}
and then obtain $\Phi_{p,L_{j,D}}$ given by \eqref{eq:EqLjD}. By substituting \eqref{eq:EqBjD} and \eqref{eq:EqLjD} into \eqref{eq:sub2YW2}, we obtain $\E[Y_jW_{j,B}]$.

Finally, we derive $\E[Y_jW_{j,U}]$ in \eqref{eq:Deltanp}. Here, we denote $Y_{j,D}$ as the time interval when $P_{n,j-1}$ and $P_{n,j}$ arrive at the computation queue at the local server at $U_n$, i.e., $Y_{j,D} = t_{j,D}-t_{j-1,D}$. Since $\E[W_{j,U}Y_j]$ is calculated as
\begin{align}\label{eq:EWjUYjcase3}
    \E[W_{j,U}Y_j] 
    &=\int_{0}^{\infty}\int_{0}^{\infty}yf_{Y_j}(y)f_{Y_{j,D}|Y_j}(y''|y)\E[W_{j,U}|Y_{j}=y,Y_{j,D}=y'']\mathrm{d}y''\mathrm{d}y,
\end{align}
we need to derive $f_{Y_{j,D}|Y_j}(y''|y)$ and $\E[W_{j,U}|Y_{j}\!=\!y,Y_{j,D}\!=\!y'']$ to obtain $\E[Y_jW_{j,U}]$. Here, $W_{j,U}$ only depends on the computation time of $P_{n,j-1}$ at the local server. Then, we obtain
\begin{align}\label{eq:WjUcase3}
    \E[W_{j,U}|Y_{j}\!=\!y,Y_{j,D}\!=\!y''] = \frac{\exp(-(\mu_n'-\lambda_n)y'')}{\mu_n'-\lambda_n}.
\end{align}

Since $B_{j,D}$ and $L_{j,D}$ are two complementary events, we rewrite $f_{Y_{j,D}|Y_j}(y''|y)$ as
\begin{align}\label{eq:fyjDincase3}
     f_{Y_{j,D}|Y_j}(y''|y)=p_{Y_{j,D},B_{j,D}|Y_j}+p_{Y_{j,D},L_{j,D}|Y_j},
\end{align}
where $p_{Y_{j,D},B_{j,D}|Y_j}$ and $p_{Y_{j,D},L_{j,D}|Y_j}$ are given by
\begin{align}
    &p_{Y_{j,D},B_{j,D}|Y_j} =  f_{Y_{j,D}|Y_j,B_{j,D}}(y''|y,B_{j,D})\mathrm{Pr}(B_{j,D}|Y_j=y)
\end{align}
and  
\begin{align}
    p_{Y_{j,D},L_{j,D}|Y_j} =  f_{Y_{j,D}|Y_j,L_{j,D}}(y''|y,L_{j,D})\mathrm{Pr}(L_{j,D}|Y_j=y),
\end{align}
respectively. 
Here, we calculate $p_{Y_{j,D},B_{j,D}|Y_j}$ as
\begin{align}\label{eq:fyjDincase3BjU}
    &p_{Y_{j,D},B_{j,D}|Y_j}
    =f_{Y_{j,D}|Y_j,B_{j,D}}(y''|y,B_{j,D})\int_{0}^{\infty}f_{Y_{j,B}|Y_j}(y'|y)
    \exp(-(\mu_D-\lambda)y')\mathrm{d}y'.
\end{align}
When $B_{j,D}$ happens, the time interval $Y_{j,D}$ depends on the transmission time of the packets generated during $Y_j$ and the transmission time of $P_{n,j}$. Note that, the PMF of the number of packets generated during $Y_j$ for other UEs is given by \eqref{eq:Ky} and the PDF of the total transmission time of $k+1$ packets is given by \eqref{eq:fYjD}. Thus, we obtain
\begin{align}\label{eq:fyjDincase3BjU1}
    f_{Y_{j,D}|Y_j,B_{j,D}}(y''|y,B_{j,D})\!&=\! \sum\limits_{k=0}^{\infty}\mathrm{Pr}(K_{j,y}\!=\!k)f_{Y_{j,D}|K_{j,y}}(y''|k)\notag\\
    &=\mu_D\exp(-\lambda_{-n}y-\mu_Dy''){I_0(2\sqrt{\lambda_{-n}\mu_Dyy''})}.
\end{align}
By substituting $f_{Y_{j,B}|Y_j}(y'|y)$ and \eqref{eq:fyjDincase3BjU1} into \eqref{eq:fyjDincase3BjU}, we obtain $p_{Y_{j,D},B_{j,D}|Y_j}$ in \eqref{eq:fyjDincase3}. We next calculate $p_{Y_{j,D},L_{j,D}|Y_j}$ in \eqref{eq:fyjDincase3} as
\begin{align}\label{eq:fyjDincase3LjU}
    f_{Y_{j,D}|Y_j,L_{j,D}}&(y''|y,L_{j,D})\mathrm{Pr}(L_{j,D}|Y_j\!=\!y)=\int_{0}^{\infty}\!f_{Y_{j\!,B}|Y_j}(y'|y)p_{Y_{j\!,D},L_{j\!,D}|Y_j\!,Y_{j\!,B}}\mathrm{d}y',
\end{align}
where
\begin{align}
    p_{Y_{j,D},L_{j,D}|Y_j,Y_{j,B}}\!&=\!f_{Y_{j\!,D}|Y_j\!,Y_{j\!,B},L_{j\!,D}}(y''|y,y',L_{j,D})\mathrm{Pr}(L_{j,D}|Y_j\!=\!y,Y_{j,B}\!=\!y').
\end{align}
Since $p_{Y_{j,D},L_{j,D}|Y_j,Y_{j,B}}$ depends on the transmission time of $P_{n,j}$ and the transmission time of packets in the transmission queue when $P_{n,j}$ arrives at the transmission queue, we obtain
\begin{align}\label{eq:fYjBLcase31}
    p_{Y_{j,D},L_{j,D}|Y_j,Y_{j,B}}
    &=\frac{(\mu_D-\lambda)(\mu_D-\lambda_{-n})}{(2\mu_D-\lambda-\lambda_{-n})}\big(\exp(-(\mu_D-\lambda)(y'-y''))\notag\\
    &-\exp(-(\mu_D-\lambda)y'-(\mu_D-\lambda_{-n})y'')\big),
\end{align}
for $y''<y'$, and 
\begin{align}\label{eq:fYjBLcase32}
     p_{Y_{j,D},L_{j,D}|Y_j,Y_{j,B}}
    &=\frac{(\mu_D-\lambda)(\mu_D-\lambda_{-n})}{(2\mu_D-\lambda-\lambda_{-n})}\big(\exp(-(\mu_D-\lambda_{-n})(y''-y'))\notag\\
    &-\exp(-(\mu_D-\lambda)y'-(\mu_D-\lambda_{-n})y'')\big),
\end{align}
for $y''\geq y'$. By substituting $f_{Y_{j,B}|Y_j}(y'|y)$, \eqref{eq:fYjBLcase31}, and \eqref{eq:fYjBLcase32} into \eqref{eq:fyjDincase3LjU}, we obtain $p_{Y_{j,D},B_{j,D}|Y_j}$ in \eqref{eq:fyjDincase3}. In addition, by substituting \eqref{eq:WjUcase3} and \eqref{eq:fyjDincase3} into \eqref{eq:EWjUYjcase3}, we obtain
\begin{align}\label{eq:E3YjWj3}
    \E[W_{j,U}Y_j] = \Phi_{p,B_{j,U}}+\Phi_{p,L_{j,U}},
\end{align}
where $\Phi_{p,B_{j,U}}$ is given by \eqref{eq:EqBjU} and $\Phi_{p,L_{j,U}}$ is given by \eqref{eq:EqLjU}. Finally, we substitute \eqref{E1YjWj1}, \eqref{eq:sub2YW2}, and \eqref{eq:E3YjWj3} into \eqref{eq:Deltanp} and obtain the average AoI of $U_n$, which is given in \eqref{eq:expreAoIpar}.

We then drive the average PAoI. Based on \eqref{eq:newPAoIevox}, the average AoI of $U_n$ in the partial computing scheme is calculated as
\begin{align}\label{eq:Omeganp}
     \Omega_{n,p}=&\frac{1}{\lambda_n}+\frac{1}{\mu_B'}+\frac{1}{\mu_D}+\frac{1}{\mu_n'}+\E[W_{j,B}]+\E[W_{j,D}]+\E[W_{j,U}].
\end{align}
According to Lemma \ref{Lemma2}, the PMF of the number of packets in the computation queue at the edge server when $P_{n,j}$ is generated  is given by
\begin{align}
    \mathrm{Pr}(K_{j,e}=k) = \left(\frac{\lambda}{\mu_B'}\right)^k\left(1\!-\!\frac{\lambda}{\mu_B'}\right).
\end{align}
Hence, we obtain
\begin{align}\label{eq:PAoIWjBPar}
    \E[W_{j,B}] = \E[K_{j,e}]\E[S_{D}]= \frac{\lambda}{\mu_B'(\mu_B'-\lambda)}.
\end{align}
Similarly, we calculate $\E[W_{j,D}]$ and $\E[W_{j,U}]$ by
\begin{align}\label{eq:PAoIWjDPar}
    \E[W_{j,D}] = \frac{\lambda}{\mu_D(\mu_D-\lambda)},
\end{align}
and
\begin{align}\label{eq:PAoIWjUPar}
    \E[W_{j,U}] = \frac{\lambda_n}{\mu_n'(\mu_n'-\lambda_n)}.
\end{align}
By substituting \eqref{eq:PAoIWjBPar}, \eqref{eq:PAoIWjDPar}, and \eqref{eq:PAoIWjUPar} into \eqref{eq:newPAoIevox}, we obtain the average PAoI of $U_n$, which is given in \eqref{eq:exprePAoIpar}.

\section{Proof for Theorem \ref{Theorem6}}\label{Appendix:B}

We first derive the upper bound on the average AoI. Based on \eqref{eq:newAoIevox}, we find that $W_{j,B}$, $W_{j,D}$, and $W_{j,U}$ are negatively correlated with $Y_{j}$, respectively, i.e., $\E[W_{j,B}Y_{j}]<\E[W_{j,B}]\E[Y_{j}]$,
$\E[W_{j,D}Y_{j}]<\E[W_{j,D}]\E[Y_{j}]$,
and 
$\E[W_{j,U}Y_{j}]<\E[W_{j,U}]\E[Y_{j}]$.
Hence, we obtain that
\begin{align}
    \Delta_n \!<\!  \frac{1}{\lambda_n}\!+\!\E[S_{j,B}]\!+\!\E[S_{j,D}]\!+\!\E[S_{j,U}]\!+\!\lambda_n\left(\E[Y_j]\E[W_{j,B}]\!+\!\E[Y_j]\E[W_{j,D}]\!+\!\E[Y_j]\E[W_{j,U}]\right)
     \!=\!\Omega_n.
\end{align}

We then derive the lower bound on the average AoI. Based on \eqref{E1YjWj1}, we obtain
\begin{align}\label{eq:AoIlower1}
    \E[Y_j W_{j,B}] &= \frac{\lambda_{-n}}{\lambda_n\mu_B'(\mu_B'-\lambda_{-n})}+\frac{\lambda_n\lambda_{-n}}{\mu_B'(\mu_B'-\lambda_{-n})^3}+\frac{\lambda_n}{(\mu_B'-\lambda_{-n})^2(\mu_B'-\lambda)}\notag\\
    &=\frac{\lambda}{\lambda_n\mu_B'(\mu_B'\!-\!\lambda)}\!+\!\frac{\lambda_n\lambda_{-n}}{\mu_B'(\mu_B'\!-\!\lambda_{-n})^3}\!-\!\frac{1}{(\mu_B'\!-\!\lambda_{-n})^2}.
\end{align}

We find that $\E[Y_jW_{j,D}]$ in \eqref{eq:sub2YW2} decreases as $\mu_B'$ increases. Hence, we obtain
\begin{align}\label{eq:AoIlower2}
   \E[Y_jW_{j,D}]\!\geq\!   \E[Y_jW_{j,D}]\big|_{\mu_B'\!\rightarrow\!\infty}\! =\! \E[Y_{j,B}W_{j,D}]\!=\!\frac{\lambda}{\lambda_n\mu_D(\mu_D\!-\!\lambda)}\!+\!\frac{\lambda_n\lambda_{-n}}{\mu_D(\mu_D\!-\!\lambda_{-n})^3}\!-\!\frac{1}{(\mu_D\!-\!\lambda_{-n})^2}.
\end{align}
Similarly, we obtain the lower bound on $\E[Y_jW_{j,U}]$, which is given by
\begin{align}\label{eq:AoIlower3}
   &\E[Y_jW_{j,U}]\geq   \E[Y_jW_{j,U}]\big|_{\mu_B'\rightarrow\infty,\mu_D\rightarrow\infty} = \E[Y_{j,D}W_{j,U}]
   =\frac{1}{\mu_n'(\mu_n'\!-\!\lambda_n)}\!-\!\frac{1}{\mu_n'^2}.
\end{align}
By substituting \eqref{eq:AoIlower1}, \eqref{eq:AoIlower2}, and \eqref{eq:AoIlower3} into \eqref{eq:newAoIevox}, and averaging $\Delta_n$ over all UEs, we obtain the lower bound on the average AoI, which is given in \eqref{eq:AoIlower}.

\section{Proof for Theorem \ref{Theorem5}}\label{Appendix:C}
By calculating the first and the second derivatives of $\Omega$ with respect to (w.r.t.) $p$, we obtain 
\begin{align}\label{eq:firstDe}
    \frac{\partial \Omega}{\partial p}=&\frac{\mu_B}{(\mu_B-p\lambda)^2}-\frac{\mu_h}{(\mu_h-(1-p)\lambda_h)^2}
\end{align}
and
\begin{align}\label{eq:d2deltaap}
    \frac{\partial^2 \Omega}{\partial p^2}=&\frac{2\lambda\mu_B}{(\mu_B-p\lambda)^3}+\frac{2(1-p)\lambda_h^2}{(\mu_h-(1-p)\lambda_h)^3},
\end{align}
respectively. From \eqref{eq:d2deltaap}, we see that $\frac{\partial^2\Omega}{\partial p^2}> 0$. It implies that there exists an optimal value of $p$ to minimize $\Omega$. 
Based on \eqref{eq:firstDe}, we obtain the optimal $p$, which is given as
\begin{align}\label{eq:optpcalcu}
    p_{opt} = \frac{\sqrt{\mu_B\mu_n}+\lambda_n-\mu_n}{\left(1+N\sqrt{\frac{\mu_n}{\mu_B}}\right)\lambda_n}.
\end{align}
By setting $p\in[0,1]$ in \eqref{eq:optpcalcu}, we obtain the optimal $p_{opt}$ given by \eqref{eq:optp}.

\end{appendices}

\bibliographystyle{IEEEtran}
\bibliography{bibli.bib}

\begin{thebibliography}{10}
\providecommand{\url}[1]{#1}
\csname url@samestyle\endcsname
\providecommand{\newblock}{\relax}
\providecommand{\bibinfo}[2]{#2}
\providecommand{\BIBentrySTDinterwordspacing}{\spaceskip=0pt\relax}
\providecommand{\BIBentryALTinterwordstretchfactor}{4}
\providecommand{\BIBentryALTinterwordspacing}{\spaceskip=\fontdimen2\font plus
\BIBentryALTinterwordstretchfactor\fontdimen3\font minus
  \fontdimen4\font\relax}
\providecommand{\BIBforeignlanguage}[2]{{%
\expandafter\ifx\csname l@#1\endcsname\relax
\typeout{** WARNING: IEEEtran.bst: No hyphenation pattern has been}%
\typeout{** loaded for the language `#1'. Using the pattern for}%
\typeout{** the default language instead.}%
\else
\language=\csname l@#1\endcsname
\fi
#2}}
\providecommand{\BIBdecl}{\relax}
\BIBdecl

\bibitem{Tang2021globecom}
Z.~{Tang}, Z.~{Sun}, N.~{Yang}, and X.~{Zhou}, ``Age of information analysis of
  multi-user mobile edge computing systems,'' in \emph{Proc. IEEE Global
  Commun. Conf.}, Madrid, Spain, Dec. 2021, pp. 1--6.

\bibitem{Simsek2016}
M.~{Simsek}, A.~{Aijaz}, M.~{Dohler}, J.~{Sachs}, and G.~{Fettweis},
  ``{5G}-enabled tactile {I}nternet,'' \emph{IEEE J. Select. Areas Commun.},
  vol.~34, no.~3, pp. 460--473, Mar. 2016.

\bibitem{Li2019}
C.~{Li}, N.~{Yang}, and S.~{Yan}, ``Optimal transmission of short-packet
  communications in multiple-input single-output systems,'' \emph{IEEE Trans.
  Veh. Technol.}, vol.~68, no.~7, pp. 7199--7203, Jul. 2019.

\bibitem{Kaul2011}
S.~{Kaul}, M.~{Gruteser}, V.~{Rai}, and J.~{Kenney}, ``Minimizing age of
  information in vehicular networks,'' in \emph{Proc. IEEE Conf. Sensor Ad Hoc
  Commun. Netw.}, Salt Lake City, UT, Jun. 2011, pp. 350--358.

\bibitem{Kaul2012}
S.~{Kaul}, R.~{Yates}, and M.~{Gruteser}, ``Real-time status: How often should
  one update?'' in \emph{Proc. IEEE Intern. Conf. Comput. Commun.}, Orlando,
  FL, Mar. 2012, pp. 2731--2735.

\bibitem{Inoue2019}
Y.~{Inoue}, H.~{Masuyama}, T.~{Takine}, and T.~{Tanaka}, ``A general formula
  for the stationary distribution of the age of information and its application
  to single-server queues,'' \emph{IEEE Trans. Inf. Theory}, vol.~65, no.~12,
  pp. 8305--8324, Dec 2019.

\bibitem{Costa2016J}
M.~{Costa}, M.~{Codreanu}, and A.~{Ephremides}, ``On the age of information in
  status update systems with packet management,'' \emph{IEEE Trans. Inf.
  Theory}, vol.~62, no.~4, pp. 1897--1910, Apr. 2016.

\bibitem{WangGC2019}
R.~Wang, Y.~Gu, H.~Chen, Y.~Li, and B.~Vucetic, ``On the age of information of
  short-packet communications with packet management,'' in \emph{Proc. IEEE
  Global Commun. Conf.}, Waikoloa, HI, Dec. 2019, pp. 1--6.

\bibitem{Yates2019}
R.~D. {Yates} and S.~K. {Kaul}, ``The age of information: Real-time status
  updating by multiple sources,'' \emph{IEEE Trans. Inf. Theory}, vol.~65,
  no.~3, pp. 1807--1827, Mar. 2019.

\bibitem{Tang2020}
Z.~{Tang}, Z.~{Sun}, N.~{Yang}, and X.~{Zhou}, ``Age of information of
  multi-source systems with packet management,'' in \emph{Proc. IEEE Intern.
  Commun. Conf.}, Dublin, Ireland, Jun. 2020, pp. 1--6.

\bibitem{moltafet2021moment}
M.~{Moltafet}, M.~{Leinonen}, and M.~{Codreanu}, ``Moment generating function
  of the {AoI} in multi-source systems with computation-intensive status
  updates,'' in \emph{Proc. IEEE Inf. Theory Workshop}, Kanazawa, Japan, Oct.
  2021, pp. 1--6.

\bibitem{zhou2021performance}
B.~Zhou and W.~Saad, ``Performance analysis of age of information in
  ultra-dense internet of things ({IoT}) systems with noisy channels,''
  \emph{IEEE Trans. Wireless Commun.}, vol.~21, no.~5, pp. 3493--3507, Nov.
  2021.

\bibitem{Tang2022Lt}
Z.~{Tang}, Z.~{Sun}, N.~{Yang}, and X.~{Zhou}, ``Whittle index-based scheduling
  policy for minimizing the cost of age of information,'' \emph{IEEE Commun.
  Lett.}, vol.~26, no.~1, pp. 54--58, Jan. 2022.

\bibitem{Hu2015White}
Y.~C. {Hu}, M.~{Patel}, D.~{Sabella}, N.~{Sprecher}, and V.~{Young}, ``Mobile
  edge computing-{A} key technology towards {5G},'' \emph{ETSI white paper},
  vol.~11, no.~11, pp. 1--16, Sep. 2015.

\bibitem{Mao2017}
Y.~{Mao}, C.~{You}, J.~{Zhang}, K.~{Huang}, and K.~B. {Letaief}, ``A survey on
  mobile edge computing: The communication perspective,'' \emph{IEEE Commun.
  Surveys Tuts.}, vol.~19, no.~4, pp. 2322--2358, Aug. 2017.

\bibitem{Mach2017}
P.~{Mach} and Z.~{Becvar}, ``Mobile edge computing: A survey on architecture
  and computation offloading,'' \emph{IEEE Commun. Surveys Tuts.}, vol.~19,
  no.~3, pp. 1628--1656, Mar. 2017.

\bibitem{Zhang2012}
{Y. Zhang}, {H. Liu}, {L. Jiao}, and {X. Fu}, ``To offload or not to offload:
  An efficient code partition algorithm for mobile cloud computing,'' in
  \emph{Proc. IEEE Intern. Conf. Cloud Netw.}, Paris, France, Nov. 2012, pp.
  80--86.

\bibitem{Chen2015}
X.~{Chen}, ``Decentralized computation offloading game for mobile cloud
  computing,'' \emph{IEEE Trans. Parallel Distrib. Syst.}, vol.~26, no.~4, pp.
  974--983, Apr. 2015.

\bibitem{Gong2013}
J.~{Gong}, S.~{Zhou}, and Z.~{Niu}, ``Optimal power allocation for energy
  harvesting and power grid coexisting wireless communication systems,''
  \emph{IEEE Trans. Commun.}, vol.~61, no.~7, pp. 3040--3049, Jun. 2013.

\bibitem{Sardellitti2015}
S.~{Sardellitti}, G.~{Scutari}, and S.~{Barbarossa}, ``Joint optimization of
  radio and computational resources for multicell mobile-edge computing,''
  \emph{IEEE Trans. Signal Inf. Process. Netw.}, vol.~1, no.~2, pp. 89--103,
  Jun. 2015.

\bibitem{Liu2016}
J.~{Liu}, Y.~{Mao}, J.~{Zhang}, and K.~B. {Letaief}, ``Delay-optimal
  computation task scheduling for mobile-edge computing systems,'' in
  \emph{Proc. IEEE Intern. Sympos. Inf. Theory}, Barcelona, Spain, Jul. 2016,
  pp. 1451--1455.

\bibitem{Mao2016}
Y.~{Mao}, J.~{Zhang}, and K.~B. {Letaief}, ``Dynamic computation offloading for
  mobile-edge computing with energy harvesting devices,'' \emph{IEEE J. Select.
  Areas Commun.}, vol.~34, no.~12, pp. 3590--3605, Sep. 2016.

\bibitem{You2017}
C.~{You}, K.~{Huang}, H.~{Chae}, and B.~{Kim}, ``Energy-efficient resource
  allocation for mobile-edge computation offloading,'' \emph{IEEE Trans.
  Wireless Commun.}, vol.~16, no.~3, pp. 1397--1411, Mar. 2017.

\bibitem{Rodrigues2018}
T.~G. {Rodrigues}, K.~{Suto}, H.~{Nishiyama}, N.~{Kato}, and K.~{Temma},
  ``Cloudlets activation scheme for scalable mobile edge computing with
  transmission power control and virtual machine migration,'' \emph{IEEE Trans.
  Comput.}, vol.~67, no.~9, pp. 1287--1300, Sep. 2018.

\bibitem{Zhao2019}
L.~{Zhao}, J.~{Wang}, J.~{Liu}, and N.~{Kato}, ``Optimal edge resource
  allocation in {IoT}-based smart cities,'' \emph{IEEE Netw.}, vol.~33, no.~2,
  pp. 30--35, Mar. 2019.

\bibitem{Ren2019}
J.~{Ren}, G.~{Yu}, Y.~{He}, and G.~Y. {Li}, ``Collaborative cloud and edge
  computing for latency minimization,'' \emph{IEEE Trans. Veh. Technol.},
  vol.~68, no.~5, pp. 5031--5044, May 2019.

\bibitem{Kuang2020}
Q.~{Kuang}, J.~{Gong}, X.~{Chen}, and X.~{Ma}, ``Analysis on
  computation-intensive status update in mobile edge computing,'' \emph{IEEE
  Trans. Veh. Technol.}, vol.~69, no.~4, pp. 4353--4366, Apr. 2020.

\bibitem{Zou2021ACM}
P.~{Zou}, O.~{Ozel}, and S.~{Subramaniam}, ``Optimizing information freshness
  through computation–transmission tradeoff and queue management in edge
  computing,'' \emph{IEEE/ACM Trans. on Netw.}, vol.~29, no.~2, pp. 949--963,
  Oct. 2021.

\bibitem{Li2021AgePr}
R.~{Li}, Q.~{Ma}, J.~{Gong}, Z.~{Zhou}, and X.~{Chen}, ``Age of processing:
  Age-driven status sampling and processing offloading for
  edge-computing-enabled real-time {IoT} applications,'' \emph{IEEE Internet
  Things J.}, vol.~8, no.~19, pp. 14\,471--14\,484, Oct. 2021.

\bibitem{Arafa2019}
A.~{Arafa}, R.~D. {Yates}, and H.~V. {Poor}, ``Timely cloud computing:
  Preemption and waiting,'' in \emph{Proc. Allerton Conf. on Commun. Control,
  and Comput.}, Monticello, IL, Sep. 2019, pp. 528--535.

\bibitem{Buyukates2022Tcom}
B.~Buyukates and S.~Ulukus, ``Timely distributed computation with stragglers,''
  \emph{IEEE Trans. Commun.}, vol.~68, no.~9, pp. 5273--5282, Sep. 2020.

\bibitem{Xu2020IoTJ}
C.~{Xu}, H.~{Yang}, X.~{Wang}, and T.~Q. {Quek}, ``Optimizing information
  freshness in computing-enabled {IoT} networks,'' \emph{IEEE Internet Things
  J.}, vol.~7, no.~2, pp. 971--985, Feb. 2020.

\bibitem{Zhou2020J}
B.~{Zhou} and W.~{Saad}, ``Minimum age of information in the internet of things
  with non-uniform status packet sizes,'' \emph{IEEE Trans. Wireless Commun.},
  vol.~19, no.~3, pp. 1933--1947, Mar. 2020.

\bibitem{wu2020data}
H.~{Wu}, H.~{Tian}, S.~{Fan}, and J.~{Ren}, ``Data age aware scheduling for
  wireless powered mobile-edge computing in industrial internet of things,''
  \emph{IEEE Trans. Industr. Inform.}, vol.~17, no.~1, pp. 398--408, Apr. 2020.

\bibitem{Wang2016T}
Y.~{Wang}, M.~{Sheng}, X.~{Wang}, L.~{Wang}, and J.~{Li}, ``Mobile-edge
  computing: Partial computation offloading using dynamic voltage scaling,''
  \emph{IEEE Trans. Commun.}, vol.~64, no.~10, pp. 4268--4282, Oct. 2016.

\bibitem{Sthapit2019}
S.~{Sthapit}, J.~{Thompson}, N.~M. {Robertson}, and J.~R. {Hopgood},
  ``Computational load balancing on the edge in absence of cloud and fog,''
  \emph{IEEE Trans. Mobile Comput.}, vol.~18, no.~7, pp. 1499--1512, Jul. 2019.

\end{thebibliography}

\end{document}